\documentclass[bibyear]{aa}
\usepackage[section]{placeins}
\usepackage{txfonts}
\usepackage{graphicx}
\usepackage{natbib}
\usepackage[usenames,dvipsnames]{color}
\definecolor{magenta}{rgb}{0.8,0,0.8}
\begin{document}

\title{Pollution versus diffusion: Abundance patterns of blue horizontal branch stars in globular clusters NGC\,6388, NGC\,6397, and NGC\,6752\thanks{Based on
    observations with the ESO Very Large Telescope at Paranal Observatory, Chile (proposal ID 69.D-0231, 69.D-0220, 65.L-0233)}
\thanks{The flux calibrated spectra, the measured equivalent widths, and the line parameters used to perform the abundance analyses are available at the CDS.}}
\author{S.\,Moehler\inst{1}
}
\institute{European Southern Observatory,
Karl-Schwarzschild-Str. 2, D 85748 Garching, Germany \email{smoehler@eso.org}
} 
\date{Received 14 October 2024 / Accepted 14 November 2024}
\titlerunning{Abundance patterns of blue HB stars in NGC\,6388, NGC\,6397 and NGC\,6752}
\abstract
{The metal-rich ([M/H] = $-$0.48) bulge globular cluster NGC\,6388
  displays a blue horizontal branch (HB). Helium enrichment, which is
  correlated with changes seen among other light elements, might
  explain this feature. Conversely, the hot HB stars in the metal-poor
  globular clusters NGC\,6397 and NGC\,6752 display high abundances of
  heavy elements caused by radiative levitation.  }
{I want to determine the abundances of cool blue HB stars in NGC\,6388
  to verify whether they are helium-enriched. To exclude the effects
  of radiative levitation for NGC\,6388 and to investigate the
  abundance changes caused by radiative levitation, I analysed the
  blue HB stars in NGC\,6397 and NGC\,6752.}
{I obtained high-resolution spectra for all three clusters. I then
  determined the effective temperatures and surface gravities from
  UV-optical photometry (NGC\,6388) and the spectra (NGC\,6397,
  NGC\,6752) together with local thermodynamic equilibrium (LTE) model
  spectra.  The results were used together with equivalent widths by
  the GALA program, which provides consistent atmospheric parameters
  and abundances.}
{For NGC\,6397 and NGC\,6752, only moderately hot HB stars were
      suitable for analysis with GALA. When including the literature data,
       a large scatter is seen at the onset of radiative
      levitation, followed by increasing abundances up to about
      13\,500\,K (Si, Fe), then turning to a plateau (Si) and a forking
      (Fe) for higher temperatures.  In NGC\,6388, the star 4113 shows
      variations in radial velocity, which may indicate binarity. For
      the remaining three blue HB stars, the metal abundances are
      consistent with the products of hot hydrogen burning. The data
      were too noisy to allow for the  helium abundances to be measured. }
{The presence of hot hydrogen burning products in the blue HB
      stars in NGC\,6388 could indicate helium enrichment. The
      abundance variations with temperature in moderately hot HB stars
      in NGC\,6397 and NGC\,6752 suggest an influence of parameters
      beyond rotation and effective temperature.}

\maketitle
\section{Introduction}
\label{sec:intro}

Low-mass stars that burn helium in a core of about 0.5\,$M_\sun$ and
hydrogen in a shell populate a roughly horizontal region in the
optical colour-magnitude diagrams of globular clusters, which has
earned them the name horizontal branch (HB) stars
\citep{ten27}. The relative distribution of stars along
the horizontal branch, namely, from red to blue, is determined to first order  by
the metallicity of the parent globular cluster. Specifically, more metal-rich
globular clusters have predominantly red horizontal branches,
while more metal-poor globular clusters have
more blue horizontal branch stars.  However, it had been noticed early on
\citep{sawi67,vdbe67} that metallicity alone cannot describe the full
variety of horizontal branch morphologies observed in Galactic
globular clusters. Much work has been dedicated in the past decades to
identify the second parameter responsible for the variety and the
excellent overview by \citet{Gratton+2010} points out that most likely
several additional parameters play important roles.
 I discuss the two processes that might affect the chemical abundances
observed for blue HB stars, namely pollution and diffusion, in this paper.

\citet{Bedin+2004} discovered that the main sequence of 
 the massive globular cluster $\omega$ Centauri is split into two
sequences. \cite{Norris2004} had first suggested helium enrichment as
a possible explanation for this split. Spectroscopic observations by
\citet{Piotto+2005} showed that (contrary to expectation)  the
stars on the blue main sequence were more metal-rich than those on the
redder dominant main sequence. \citet{Piotto+2005}
also found tentative evidence that
the bluer stars show strong enrichment in nitrogen and that a helium
enrichment up to $Y$=0.35\ldots0.45 is required to reproduce the bluer
main sequence. Such abundance changes cannot happen in low-mass main
sequence stars, so the material from which the stars formed must have
been polluted. Unfortunately helium cannot be observed directly in
main sequence stars of Galactic globular clusters. It can be observed
in cool blue horizontal branch stars with effective temperatures
between about 8\,500\,K (when stars are hot enough to show helium
absorption in their spectra) and 11\,500\,K because hotter horizontal
branch stars are affected by diffusion (see below for more
information).

Since 2004 multiple populations have been detected in most Galactic
globular clusters that have sufficiently good photometric data and for
many also spectroscopic data are available. \cite{ARAA2018} provided an
excellent review of the situation. For my analysis their summary
of abundance variations expected to accompany helium enrichment is of
special importance: For stars expected to be enriched in helium,  we would
also expect enrichment of nitrogen and sodium, accompanied by
the depletion of oxygen and carbon. The enrichment of magnesium and depletion
of aluminium are less clearly observed. On the other
hand, \cite{Vaca+2024}  experimented with various pollution models and ultimately demonstrated that only
the abundances of aluminium and helium show a clear correlation,
whereas the abundances of nitrogen and sodium appear decoupled
from those of helium.

The effects of diffusion in blue HB stars, instead, are limited to the
atmospheres of the stars. A large photometric survey of many globular
clusters by \citet{Grundahl+99} demonstrated that the Str\"omgren
$u$-brightness of blue HB stars suddenly increases near 11\,500\,K.
This $u$-jump (the so-called `Grundahl jump') is attributed to a
sudden increase in the atmospheric metallicity of the blue HB stars to
super-solar values that is caused by the radiative levitation of heavy
elements.  \citet{beco99,beco00} and \citet{mosw00} confirmed with
direct spectroscopic evidence that the atmospheric metallicity does
indeed increase to solar or super-solar values for HB stars hotter
than the $u$-jump. These findings also helped to understand the cause
of the low-gravity problem: \citet{crro88} and \citet{mohe95,mohe97}
found that hot horizontal branch stars (when analysed with model
spectra of the same metallicity as their parent globular cluster) 
show significantly lower surface gravities than expected from
evolutionary tracks. Analysing them  with more appropriate
metal-rich model spectra instead reduces the discrepancies considerably
\citep{mosw00}. The more realistic stratified model atmospheres of
\citet{hulh00} and \citet{lmhh09} reduce the discrepancies in surface
gravity even more (see \citealt{Moehler+2014} for more details).
Along with the enhancement of heavy metals, a decrease in the helium
abundance is observed for stars hotter than $\approx$11\,500\,K, while
cooler stars have normal helium abundances within the observational
errors.
 I describe in brief the globular clusters studied here below.

NGC\,6388 is a massive, metal-rich globular cluster in the
Galactic bulge.  Using \textit{Hubble} Space Telescope observations,
\citet{riso97} found an unexpected population of hot HB stars in
this cluster and NGC\,6441. This makes these clusters the most metal-rich
clusters to show the 'second parameter' effect.  Ordinarily,
metal-rich globular clusters have only a red HB clump, whereas
NGC\,6388 and NGC\,6441 show extended blue HB tails containing
${\approx}$15\% of the total HB population. \cite{Busso+2007} find
that the horizontal branch in NGC\,6388 even extends to the very
hottest HB stars, namely, the so-called blue hook stars.

Helium-enriched populations could explain these findings, because such
populations have (for a given age)  lower mass stars at the tip of
the red giant branch; namely, immediately before the onset of helium-core
burning. Because the core mass of HB stars is roughly constant at
0.45\,M$_\odot$ to 0.50\,M$_\odot$, lower mass main sequence stars
result in HB stars with smaller hydrogen envelopes and thus higher
effective temperatures.  The higher helium abundance also increases
the luminosity of the HB stars, due to higher energy output of the
hydrogen shell burning process. This has been illustrated, for instance, by
\citet{Lee+2005}. This might explain the tilt of the horizontal
branch in NGC\,6388, which shows higher luminosities for
bluer HB stars. \cite{Busso+2007} find that a range of helium abundances,
$Y$, from 0.26 to 0.38 is required to explain the full extent of the
horizontal branch in NGC\,6388 and that differential reddening cannot
explain the observed HB tilt.

For NGC\,6388 and NGC\,6441, \citet{Brown+2016} find that the Grundahl
jump is shifted to higher effective temperatures of about 13\,500\,K
to 14\,000\,K, which is attributed to the large extent of helium enhancement
required to explain the existence of blue HB stars in these metal-rich
globular clusters. I observed blue HB stars in NGC\,6388 to verify if they show signs of pollution like helium enrichment or abundance patterns consistent with products of hot hydrogen burning.

NGC\,6397 is a very nearby metal-poor globular cluster ([Fe/H] =
$-$1.95, \citealt{Harris96}, December 2010 version), which displays a short
blue horizontal branch. NGC\,6752 is another nearby globular
cluster of intermediate metallicity ([Fe/H] = $-$1.56,
\citealt{Harris96}, December 2010 version), displaying a blue horizontal
branch extending to high effective temperatures. 
For both clusters, only small enrichment in helium is expected
\citep{Marino+2014}, which is consistent with the results of 
\citet{Villanova+2009} from observations of cool blue HB stars in
NGC\,6752.

Blue HB stars in
these two globular clusters were observed with the goal to study
abundance changes around the Grundahl jump. In addition, the blue HB stars in NGC\,6752 cover a larger range in temperature, which allows for an investigation of abundance variations with the effective temperature.
These data are analysed here together with the spectra of the blue HB
stars in NGC\,6388 to ensure that any possible effects of diffusion (although this is not expected at the effective temperatures of the HB stars in
NGC\,6388) can be distinguished from the effects of pollution.
Section\,\ref{sec:obs_proc} and Appendices \ref{sec:app_obs_info} and \ref{app_ngc6388_obs}  describe the observations and their reduction, with the analysis following in Sects.\,\ref{sec:params} and \ref{sec:abu}. The results are discussed in Sect.\,\ref{sec:discussion}.

\section{Observations and data reduction}\label{sec:obs_proc}
All observations described in this section were taken with the UV-Visual Echelle Spectrograph (UVES) at the European Southern Observatory's (ESO's)
Very Large Telescope (VLT) Unit Telescope 2 in the
  years 2000 and 2002. The globular clusters NGC\,6388 (69.D-0231(B))
  and NGC\,6397 (69.D-0220(B)) were observed only in 2002, while
  NGC\,6752 was observed in 2000 (65.L-0233) and 2002
  (69.D-0220(A)). Below, I describe the observations grouped by
  observing run, because all data from a given run were observed in
  the same year and with the same instrumental setup. Detailed
information on the targets (coordinates, observing times and
conditions, photometric information) can be found in
Appendix\,\ref{sec:app_obs_info}.

\subsection{NGC\,6388 (69.D-0231(B), 2002)}
The spectroscopic targets in NGC\,6388 were selected from the catalogue of
\citet{Piotto+97}, with an aim to select the most isolated stars from the
original WFPC2 images, obtained close to the core of the
globular cluster. With 69.D-0231(A) a larger sample of blue HB stars
in NGC\,6388 was observed with the FOcal Reducer/low dispersion Spectrograph 2 (FORS2) at low spectral
resolution. The results of those observations together with finding
charts for all stars are reported in \citet{Moehler+06}. Information on the targets discussed here is provided in Table\,\ref{tab:targobs}.

  The UVES
observations used the dichroic mode with DICHR\#1, slit width of
1\farcs0 and the settings 390\,nm and 564\,nm. These cover the
wavelength ranges 326\,nm to 454\,nm (BLUE), 458\,nm to 559\,nm (REDL), and 569\,nm to 668\,nm (REDU). They should provide a resolution
$\lambda/\Delta\lambda$ of about 47500, 43500, and 43000 for BLUE,
REDL, and REDU data, respectively, assuming that the star fills the
slit. At the centre of the spectra this corresponds to a full width at
half maximum (FWHM) of 6.3\,pixels, 9\,pixels, and 11\,pixels of
0.013\,\AA\ for BLUE, REDL, and REDU spectra, respectively.
All spectra were processed using the ESO UVES pipeline (uves-6.1.4)
together with the Esoreflex workflow \citep{Reflex}. Instead of the
default pipeline parameters, I used a $\kappa$ value of 4 for cosmic
ray rejection during optimal extraction and fixed wavelength steps of
0.013\,\AA\ for all three arms. When inspecting the results, several
problems with data observed in the red arm were discovered and these are
described in detail in Appendix\,\ref{app_ngc6388_obs}.

The extracted flux-calibrated spectra were then corrected to
barycentric velocities and scaled in flux to the first spectrum
observed for a given target and setup, using the median values of the
flux in the intervals: 4200\,\AA--4220\,\AA, 4720\,\AA--4740\,\AA, and
6200\,\AA--6220\,\AA. These scaled spectra were then averaged,
clipping flux values deviating by more than
$\pm 5\cdot10^{-17}$\,erg\,s$^{-1}$\,cm$^{-2}\,\AA^{-1}$ from
the median.

From the stacked spectra, I then determined radial velocities using the
cores of H$\delta$ (BLUE), H$\beta$ (REDL), and H$\alpha$ (REDU),
respectively. This resulted in average radial velocities of
$+86.6\pm0.6$\,km s$^{-1}$ for star 1233, $+75.2\pm0.4$\,km
s$^{-1}$ for star 5235, and $+85.1\pm1.0$\,km s$^{-1}$ for
  star 7788, compared to radial velocities of the cluster of 81.2\,km
s$^{-1}$ \citep{Carretta+2022} and 82.5\,km s$^{-1}$
\citep{Carretta+2023}. The stars
discussed here are rather close to the centre of NGC\,6388 (using
the coordinates from \citealt{Vasiliev+2021}) with distances of about
150\arcsec\ (4113, 5235, 7788) and 180\arcsec\ (1233), where Fig. 1
of \citet{Carretta+2023} shows a scatter in radial velocity of about
$\pm$10\,km \,s$^{-1}$.

After applying the barycentric corrections the spectra of the star
4113 showed variations in line positions between subsequent exposures,
which had all been processed with the same flat field and wavelength
calibration data. To see whether these variations correspond to radial
velocity variations or, instead, to constant wavelength shifts, I
cross-correlated short chunks of the spectra ($\le$10\,\AA), using the
first exposure as reference. I found average relative radial
velocities for the BLUE spectra of $-5.2\pm2.4$\,km s$^{-1}$,
$+10.5\pm2.3$\,km s$^{-1}$, and $+14.7\pm5.3$\,km s$^{-1}$, for the
second, third, and fourth exposure, respectively. For the REDL spectra,
the corresponding values are $-5.0\pm1.3$\,km s$^{-1}$,
$+11.2\pm2.6$\,km s$^{-1}$, and $+13.1\pm1.7$\,km s$^{-1}$. The fact that
the velocities between the two independently calibrated arms agree
points towards real changes in radial velocities as opposed to
differences in the wavelength calibration. The large rms for F275W and
F438W in the \textit{Hubble} Space Telescope UV Legacy Survey of Galactic
Globular Clusters (HUGS) data (see Table\,\ref{tab:HUGS}) also points
towards some kind of variability. The REDU data were unfortunately too
noisy to permit reliable measurements. I applied the velocity
corrections to the individual spectra and stacked them similarly to
the other ones, assigning weight of 1 to the first 2 spectra and 0.5
to the last two spectra, which have substantially lower
signal-to-noise ratios.  Daospec \citep{daospec} found an average
velocity of $+81.3\pm0.8$\,km s$^{-1}$ for the stacked spectrum, again
in good agreement with the globular cluster velocity.


\subsection{NGC\,6397 and NGC\,6752 (69.D-0220, 2002)}
The spectroscopic targets in NGC\,6397 (69.D-0220(B)) were selected from the catalogue
of \citet{Twarog2000},
while those in NGC\,6752 (69.D-0220(A)) were selected from the catalogue of
\citet{Buonanno+86} and cross-matched with the catalogue of
\citet{Grundahl+99}. In both clusters stars close to the Grundahl gap,
where radiative levitation sets in, were selected. The same instrument
and pipeline parameter settings were used as for NGC\,6388.
Information on the coordinates and observing times and conditions is given in
Tables\,\ref{tab:targobs6397} (NGC\,6397) and \ref{tab:targobs6752} (NGC\,6752).


The extracted flux-calibrated spectra were then corrected to
barycentric velocities. At the same time the spectra of the stars in
NGC\,6752 and NGC\,6397 were corrected for the cluster velocities of
$-$26.7\,km\,s$^{-1}$ \citep{Lardo+15} and $+$17.8\,km\,s$^{-1}$
\citep{Husser+16}, respectively.
Next, the spectra were scaled in flux to the first spectrum
observed for a given target and setup, using the median values of the
flux in the intervals 4280\,\AA--4290\,\AA, 4980\,\AA--4990\,\AA, and
6080\,\AA--6090\,\AA. 

The spectra obtained using optimal extraction show in several cases
strong spikes, which are artefacts of the extraction algorithm,
because they are not seen in one-dimensional spectra calculated as an
unweighted sum. Because the overall quality of the optimally extracted
spectra is much better than that of unweighted sum, I chose to
correct for the spikes instead of using the sum.

The spikes in the BLUE data for NGC\,6752 from 13 June 2002 and
31 July 2002 were fortunately sufficiently shifted between exposures due
to the earth's motion so that they did not overlap in the spectra from
the two dates. I used the average flux from the unaffected scaled
spectra to replace the flux in the combined spectra. Only B3243 showed
some spikes in its REDL spectra, which could be fixed in the same way.
  For B652 and B944 I had only spectra from one date. There I
  smoothed the spectra with a median filter of 21 pixels width,
  replacing only those pixels whose filtered flux differed from their
  original flux by more than 20\%.

 The BLUE data for NGC\,6397 from all observing dates showed extreme spikes,
  which were mostly present in both spectra of a given target. Because
  I had only spectra from one date per star, I smoothed the spectra
  with a median filter of 37 pixels width, replacing only pixels whose
  filtered flux differed from their original flux by more than
  20\%. The flux from these smoothed spectra was then used to replace
  the regions affected by the spikes, which were
fortunately located at wavelengths far from
  any significant spectral features. In those cases where the spikes
  appeared only in one of the two spectra I replaced the flux in the
  combined spectrum with the flux from the unaffected spectrum.
After these corrections, the scaled spectra were averaged per star.

\begin{table*}[ht!]
  \caption[]{Photometric data and membership probabilities for the target stars in NGC\,6388 from the HUGS survey (\citealt{Piotto+HUGS},
\citealt{Nardiello+HUGS}), derived with Method1. Only the filter 275W has more than one observation and no rms is given for the filters F606W and F814W.  }\label{tab:HUGS}
  \begin{tabular}{llrrllllllllll}
    \hline
    \hline
    star & HUGS ID & X & Y & \multicolumn{3}{c}{F275W}& \multicolumn{2}{c}{F336W}& \multicolumn{2}{c}{F438W}& \multicolumn{1}{c}{F606W}& \multicolumn{1}{c}{F814W} & Memb\\
    & & &                 & brightn. & rms  & n&     & rms   &    & rms   &    &   & \\
    & & [pixel] & [pixel] & [mag] & [mag] & & [mag] & [mag] & [mag] & [mag] & [mag] & [mag & [\%]\\
    \hline
1233 & R0021699 & 3016.1 & 6328.2 & 19.086 & 0.013 & 2 & 17.982 & 0.026 & 17.308 & 0.008 & 16.700  & 16.169  & 97.4\\
4113 & R0007211 & 3473.8 & 4571.5 & 19.696 & 0.376 & 2 & 18.455 & 0.001 & 17.566 & 0.345 & 16.945  & 16.089  & 94.3\\
5235 & R0001391 & 3903.2 & 3323.9 & 18.744 & 0.017 & 2 & 17.869 & 0.010 & 17.426 & 0.011 & 16.955  & 16.529  & 97.8\\
7788 & R0022757 & 3752.3 & 6825.7 & 18.793 & 0.018 & 1 & 17.994 & 0.012 & 17.686 & 0.005 & 17.209  & 16.806  & 97.8\\
    \hline
  \end{tabular}
\end{table*}
\subsection{NGC\,6752 (65.L-0233(A), 2000)}
The spectroscopic targets in NGC\,6752 were selected from the catalogue of
\citet{Buonanno+86} to cover the full range of the blue and extreme horizontal branch in this cluster.
Information on the coordinates and observing times and conditions is given in
Table\,\ref{tab:targobs6752}.

The observations used only the BLUE arm with the setting 437\,nm,
which covers the wavelength ranges 373\,nm to 499\,nm, and a slit
width of 1\farcs2, which should provide a resolution,
$\lambda/\Delta\lambda$, of about 39600, assuming that the star fills
the slit. At the centre of the spectra this corresponds to an FWHM of 8.5\,pixels of 0.013\,\AA. However, about 70\% of
the observations had a seeing better than 1\farcs2 (taking the
observed airmass into account), resulting in a higher
resolution. Therefore, the same spectral
resolution will be used as for the
observations described previously.
The same pipeline parameter settings were used as for NGC\,6388.

\section{Atmospheric parameters}\label{sec:params}
Any abundance analysis  requires at least starting values for effective
  temperatures and surface gravities, which may then be refined during the
  abundance analysis. Below, I describe how those starting values were
  determined.

  \subsection{NGC\,6388}\label{ssec:NGC6388_par}

To get a first estimate of the atmospheric parameters effective
temperature and surface gravity, I used the NGC\,6388
observations from the \textit{Hubble} Space Telescope UV Globular Cluster Survey (HUGS,
\citealt{Piotto+HUGS}, \citealt{Nardiello+HUGS})\footnote{\url{http://dx.doi.org/10.17909/T9810F}}
Because
the stars discussed here are bright enough to be measurable in
individual exposures, I used the magnitudes listed in
  Table\,\ref{tab:HUGS}; these values were determined with Method 1 (see
\citealt{Nardiello+HUGS} for details).  I converted the observed
photometric data into flux values using the Vega model
alpha\_lyr\_mod\_004.fits provided at the CALSPEC
page of the Space Telescope Science Institute\footnote{\url{https://www.stsci.edu/hst/instrumentation/reference-data-for-calibration-and-tools/astronomical-catalogs/calspec}}
\citep,{Bohlin+14,
  Bohlin+20} together with the
Wide Field camera 3 filter throughputs\footnote{\url{https://www.stsci.edu/files/live/sites/www/files/home/hst/instrumentation/wfc3/performance/throughputs/\_documents/UVIS.zip}}.
I divided for each filter the
integrated flux of the Vega spectrum multiplied with the throughput
curve of the filter by the integrated throughput curve. I used the
results of this exercise as flux values for magnitude=0. Next, I used
ATLAS9 flux spectra for [M/H] = $-$0.5 provided in the
CD-ROM No. 13 of the {ATLAS Stellar
    Atmospheres Program and 2 km s$^{-1}$ grid}\footnote{\url{http://kurucz.harvard.edu/}}. These
model spectra were reddened using the
python routines\footnote{\url{https://dust-extinction.readthedocs.io/en/stable/}}
provided by \cite{Fitzpatrick+19} and scaled to the flux
observed in the F814W filter. Because of the variable reddening in
NGC\,6388, I used values of 0.5 and 0.45 for $E_{B-V}$. Then the best
fitting model spectrum was determined by eye, plotting the scaled and
reddened model spectrum on top of the HUGS photometric data. The
results are given in Table\,\ref{ngc6388_photpar}.

\begin{table}[ht]
  \caption[]{Effective temperature and surface gravity estimates for the HB stars in NGC\,6388
    obtained from HUGS (\citealt{Piotto+HUGS},
\citealt{Nardiello+HUGS}) data, using Kurucz model spectra for [M/H] =
    $-$0.5. The values used for further analysis are marked in {\bf bold font}}\label{ngc6388_photpar}
  \begin{tabular}{llcc}
    \hline
    \hline
    star & $E_{B-V}$ & $T_{eff}$ & $\log g$\\
& [mag]    & [K]       & [cgs]\\
    \hline
    1233 & 0.45 & 8250\ldots{\bf 8500} & 3.0\ldots{\bf 3.5}\\
    & 0.50 & 8750\ldots9000 & 3.5\\[2mm]
    4113   & 0.45 & 6500 & 3.5 \\
    & 0.50 & {\bf 6750} & {\bf 3.5}\\[2mm]
    5235 & 0.45 & {\bf 9500}\ldots9750 & {\bf 3.5}\ldots4.0\\
    & 0.50 & \multicolumn{2}{c}{no fit}\\[2mm]
    7788 & 0.45 & {\bf 10000}\ldots10500 & {\bf 4.0}\ldots4.5\\
    7788 & 0.50 & 10500\ldots11000 & 4.0\ldots4.5 \\ 
    \hline
    \end{tabular}
\end{table}
\FloatBarrier
\subsection{NGC\,6397 and NGC\,6752}\label{ssec:others_par}
To determine the atmospheric parameters effective temperature and
surface gravity, I fitted  local thermodynamic equilibrium (LTE) model spectra of various metallicities
containing only hydrogen and helium lines (HHe model
    spectra) to the observed flux-calibrated spectra.

 I checked the wavelength range 4580\,\AA--4680\,\AA\ for metal lines,
 which indicate the presence of radiative levitation in hot HB stars
 in metal-poor globular clusters. For stars with many lines, I used HHe model spectra of solar metallicity \citep{mosw00} and also fit 
 the helium abundance, using the He\,{\sc i} lines at 4026\,\AA,
 4388\,\AA, and 4471\,\AA. 

Because the model spectra do not include metal lines, I tried to mask
at least the strongest observed ones. For that purpose I
searched for narrow lines with a depth of about 10\% to 20\% of the
continuum level, depending on the signal-to-noise ratio of the
spectra. The lines that were found in this way were masked $\pm$0.26\,\AA\ (20 pixels)
around each line during the fit process to avoid systematic errors in
the continuum definition and the actual line profile fit. Obviously,
this procedure will not remove faint metal lines.

For stars with few lines I used instead HHe model spectra with
solar helium abundance and metallicities of $-$1.5 (NGC\,6752,
\citealt{mosw00}) and [M/H] = $-$2.0 (NGC\,6397, calculated in the
same way as the ones in \citealt{mosw00}). For these stars I did not
try to fit the helium abundance.

To establish the best fit to the observed spectra, I used the routines
developed by \citet{Bergeron+92} and \citet{Saffer+94}, as modified by
\citet{Napiwotzki+99}, which employ a $\chi^2$ test. The $\sigma$
necessary for the calculation of $\chi^2$ is estimated from the noise
in the continuum regions of the spectra. The fit program normalises
modelled spectra and observed spectra using the same points for the
continuum definition. I fitted the Balmer lines H12 to H$\beta$
(excluding H$\epsilon$ due to the interstellar Ca\,{\sc ii} absorption
line and H$\alpha$ due to the low signal-to-noise ratio in the REDU
arm). These fit routines underestimate the formal errors by at
least a factor of 2 (Napiwotzki priv. comm.). Therefore, I provide
the formal errors multiplied by 2 to account for this effect. In addition,
the errors provided by the fit routine do not include possible
systematic errors due to, for instance, flat field inaccuracies or imperfect
sky subtraction.
The results of the fitting process can be found in Tables\,\ref{tab:metal_poor_param} and \ref{tab:metal_poor_param_diff}.
\section{Abundances}\label{sec:abu}

Because
the metal lines are well resolved and rarely blended, I decided to use
equivalent widths instead of spectrum synthesis to determine
abundances to get a better handle on the uncertainties.

The flux-calibrated spectra were divided by the best fitting model
spectra (normalised to a continuum of 1) to remove the strong hydrogen
absorption lines. To keep the helium lines in the observed spectra, I
interpolated the model spectra across the helium lines. Next, the
resulting continuum flux of the observed spectra was fitted with a
fifth-order polynomial to normalise the continuum of the observed
  spectra to 1.

 For the stars in NGC\,6388, I followed the procedure for masking metal
 lines described in Sect.\,\ref{ssec:others_par}. I then fit HHe
 model spectra for [M/H] = $-$0.5 \citep{Moehler+06}, with a solar helium
 abundance to the observed flux-calibrated spectra. Because the line
 masking removes only strong metal lines, the results of that procedure
 are not trustworthy; however,  they do provide sufficiently good fits to the
 hydrogen lines to enable their removal from the spectra. I fit only
 the hydrogen lines H$\alpha$ to H12 with the exception of
 H$\epsilon$ because of the contamination by interstellar
 Ca\,{\sc ii} absorption lines.

\subsection{Equivalent widths}\label{ssec:equiv}
To measure equivalent widths I used Daospec \citep{daospec}.  To
ensure that the values I used for the FWHM were
reasonable, I performed the following checks:
\begin{enumerate}
\item I verified that unresolved lines were not split into two or more
  lines during the Daospec analysis, which provided a lower limit to
  the FWHM.
\item I also plotted the errors of the equivalent widths vs. the
  equivalent widths and thereby identified FWHM values which created
  systematically larger errors.
\end{enumerate}
Both criteria ruled out FWHM values below about six\,pixels, eight\,pixels,
and nine\,pixels for the BLUE, REDL, and REDU data, respectively. Finally
I used for NGC\,6388 values of 6.3 (BLUE), 9.0 (REDL), and 11.0 (REDU). For NGC\,6397 and NGC\,6752, I\ used values of 6.7 (BLUE),
8.7 (REDL), and 12.0 (REDU).

\subsection{Helium lines}\label{ssec:helium}
 He\,{\sc i} lines are a special case for Daospec, because they are
 often wider and more shallow than metal lines (see for instance in
 Fig.\,\ref{fig:rotation} the He\,{\sc i} line at 4471\,\AA\ vs the
 Mg\,{\sc ii} line at 4481\,\AA) and then split by Daospec into
 several components. Due to their larger width and shallower depth
 helium lines are also much more affected by imperfect fits of the
 continuum level. After a visual inspection, I combined all the lines
 measured by Daospec that contribute to a helium line, unless they were
 plausibly caused by some metal absorption. This resulted in
 equivalent widths with high uncertainties. Convolving the spectra
 with a Gaussian with an FWHM of 0.1\,\AA\ lead to less line splitting
 in Daospec, but did not solve the issue of the correct continuum
 definition. It only marginally improved the uncertainty of the final
 helium abundance, which is always above 0.3\,dex. Therefore, helium was
 not included in the further analysis.

\subsection{Identification of observed metal lines}\label{ssec:obs_metal}
 Daospec identifies a line by using the closest wavelength found in
 the user-provided line catalogue. Because other lines nearby may
 contribute to a measured line I ran a line identification that
 selects all lines within $\pm$0.1\,\AA\ around the rest wavelength
 derived by Daospec. If more than one line was found, I inspected the
 observed line to decide whether this was a true blend and should 
 be excluded from further analysis.

 For a start I extracted line lists from the Vienna Atomic Line
Database (VALD\footnote{\url{http://vald.astro.uu.se/~vald/php/vald.php}},
\citealt{VALD3}) for the following combinations of effective
temperature, surface gravity, and metallicity [M/H]: For stars
affected by diffusion (NGC\,6397, NGC\,6752) 15000\,K 4.5 $+$0.5,
20000\,K 4.5 $+$0.5, 25000\,K 5.0 $+$0.5; for stars unaffected by
diffusion (all clusters) 7000\,K 3.0 0.0 (4113), 9157\,K 3.07 $-$0.5 (all
others).

Later, I retrieved line lists for the parameters listed in
Tables\,\ref{tab:abu_photpar}, \ref{tab:abu_ngc6752_2000},
\ref{tab:abu_ngc6752_69D220}, and \ref{tab:abu_ngc6397} to have them
tailored to the actual atmospheric parameters. I used microturbulent
velocities of 0\,km s$^{-1}$ and requested a minimum line depth of
0.01. For the stars affected by diffusion I used solar metallicity and
for the other stars the same metallicities as for the
HHe models described earlier. For lines close to each other, I
used the information on the predicted depth of the line to decide
whether both lines affect the equivalent width. If the
predicted depth of one line was less than 5\% of that of the
other line, I removed the weaker catalogue line and kept the
stronger.

In most cases the difference between the observed wavelength of an
identified line and the value from the lines list was within
$\pm$0.03\,\AA. For 4113, however, a broad range of offsets was found
within $\pm$0.1\,\AA. Based on the experience with the more reliable
spectra of other stars, I restricted the lines used for the abundance
analysis of 4113 to those with observed wavelengths within
$\pm$0.03\,\AA\ of the wavelength from the line list. Doing so
provided a good agreement between the abundances derived from Fe\,{\sc
  i} and Fe\,{\sc ii} lines.




\subsection{Abundance analysis}\label{ssec:abu_analysis}
To derive abundances from the equivalent widths I used
GALA\footnote{\url{http://www.cosmic-lab.eu/gala/gala.php}} \citep{GALA}. This
tool performs an abundance analysis, based on the equivalent widths and
grids of model atmospheres. It simultaneously determines the
abundances, effective temperature, surface gravities, and
microturbulent velocities, together with error estimates. For more
details, see \citet{GALA}.

I changed the following parameters from the values provided in the tutorial file:
\begin{enumerate}
\item range of valid equivalent widths to cover $-9.9 \le
  \log (W_\lambda/\lambda) \le -0.1;$
  \item error in the equivalent width $\le$10\% to get robust results;
  \item ion used for optimisation: Fe\,{\sc ii} for hotter stars.
\end{enumerate}

I found that for the cool blue HB stars in NGC\,6397 (T160, T164,
T166, and T174) and in NGC\,6752 (B944, B2281, B2454, and B2735), which are
not affected by radiative levitation, neither abundances, nor
their  atmospheric parameters could be determined with GALA,
because there were too few metal lines. The same applies to the
hottest stars in NGC\,6752, namely B491, B3699, and B4548, which have
either  lines from one ionisation stage of iron only or iron lines
covering solely a narrow range of excitation energies.

For ions with few lines overall, I reran GALA with an error limit of
15\% on the equivalent widths, keeping the effective temperature, surface
gravity, and microturbulent velocity fixed to the values determined
from equivalent widths with errors of at most 10\%.

The abundances for sodium are affected by non-LTE (NLTE) effects were assumed as described in
\citet{Mashonkina+2000}. Then, I applied the following corrections to sodium
abundances determined for the blue HB stars in NGC\,6388: $-$0.66 (star 1233), $-$0.39 (star 5235), and $-$0.36 (star 7788).

The errors for the abundances are derived from the errors of the
equivalent widths and the overall scatter of the derived abundance.
Abundances are listed only if there are at least two usable lines for
the given ion and the abundance uncertainty is below 0.3\,dex.  GALA
may also take into account the uncertainties in effective temperature,
surface gravity, and microturbulent velocity; however, this often failed,
especially for the surface gravity.
Therefore, I ran GALA manually for the atmospheric
parameters$\pm$uncertainties to get an
estimate of these sources of uncertainty.  For the stars T179 and T191
in NGC\,6397, this procedure did not work because not all necessary
models converged. For details and results, we refer to
Appendix\,\ref{ssec:uncert}.
The final abundances can be found in Tables\,\ref{tab:abu_photpar}, \ref{tab:abu_ngc6752_2000},
\ref{tab:abu_ngc6752_69D220}, and \ref{tab:abu_ngc6397} in Appendix\,\ref{ssec:abu}.

\section{Discussion}\label{sec:discussion}
\subsection{Rotation}
\begin{figure}[!h]
\includegraphics[height=0.9\columnwidth,angle=270]{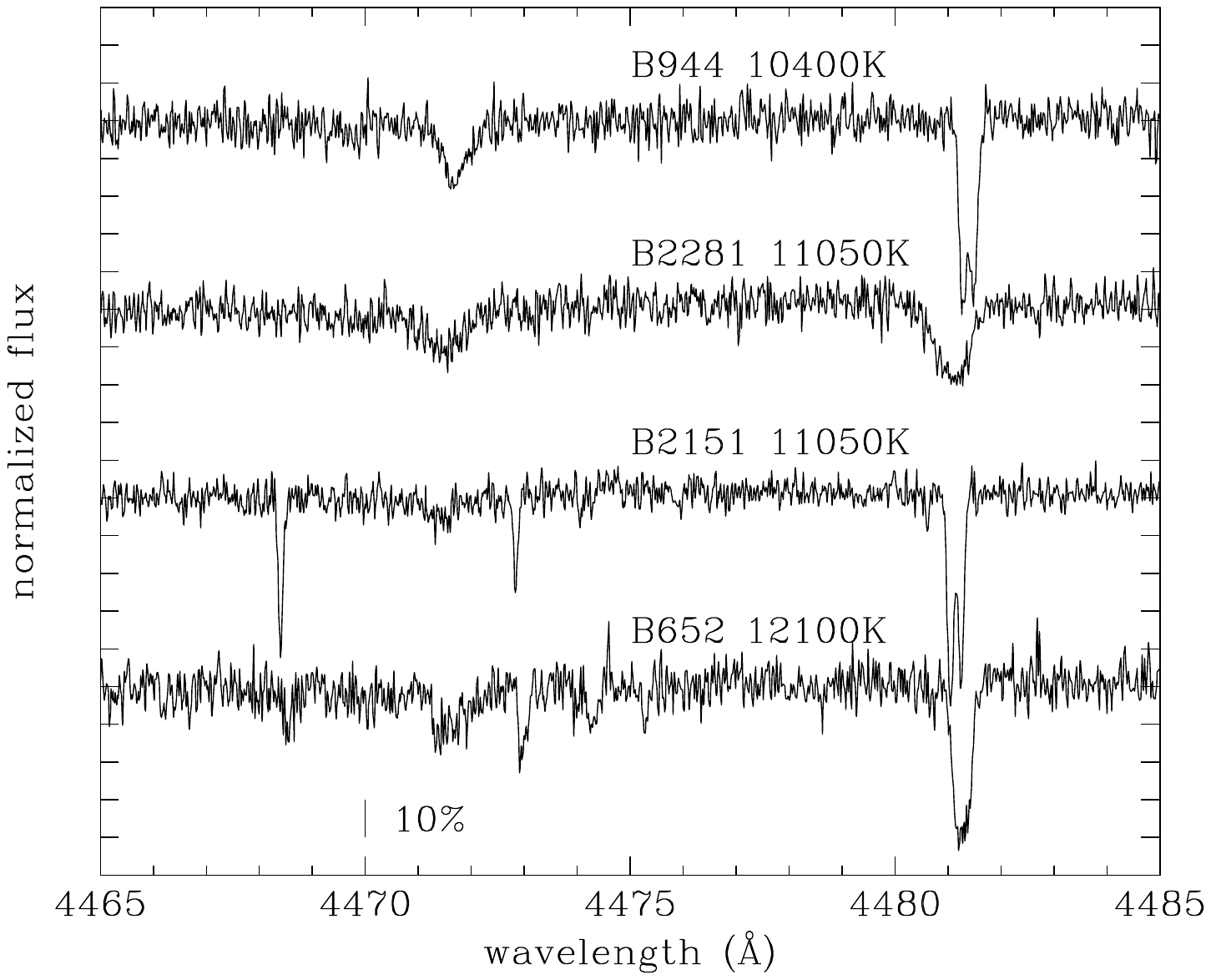}
\caption[]{Region of the He\,{\sc i} line at
  4471\,\AA\ and of the Mg\,{\sc ii} line at 4481\,\AA\ for the four
  stars in NGC\,6752. The upper two spectra belong to stars without evidence
  for diffusion and the lower two show stars with evidence for
  diffusion. The spectrum of B2281 shows clear signs of rotation,
  while the situation for B652 is less clear. The offsets in line position indicate deviations of the star's radial velocity from that of the globular cluster. }\label{fig:rotation}
\end{figure}

The stars B2281 and B652 (possibly) in NGC\,6752 show signs
rotational broadening in their spectra (see
Fig.\,\ref{fig:rotation}). This is surprising for B652, because the star shows evidence for diffusion, which usually coincides with
negligible rotational velocities. The fact that the He\,{\sc i} line at
4471\,\AA\ is visible in the spectrum of B652 suggests that the
diffusion is limited in this star.

\subsection{Evidence for pollution in NGC\,6388}\label{ssec:rel_abu_6388}

\begin{figure}[!h]
\includegraphics[height=\columnwidth,angle=270]{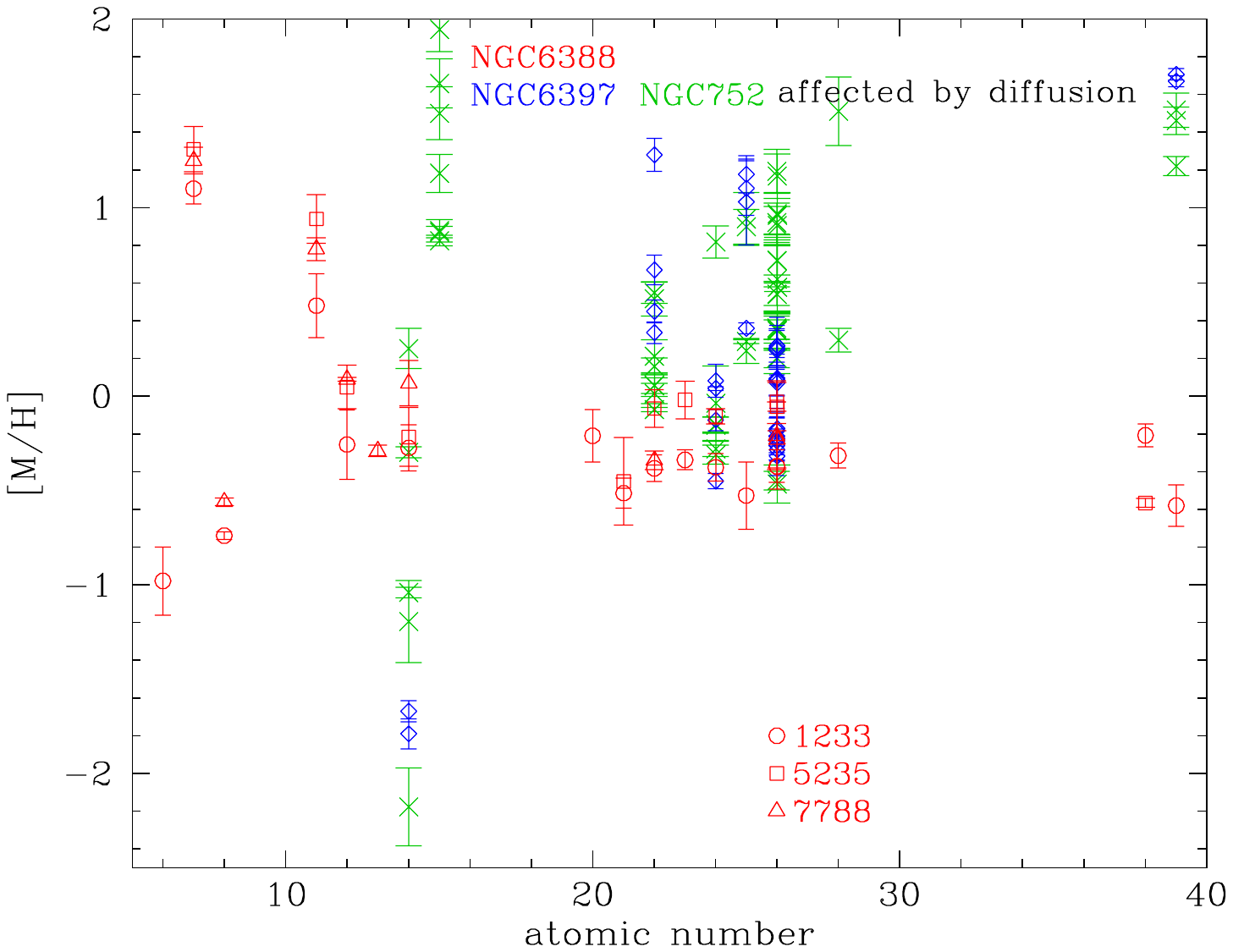}
\caption[]{Abundances of stars in NGC\,6388 (red symbols) and for stars affected by diffusion in
  NGC\,6397 (blue lozenges) and NGC\,6752 (green crosses) relative to the solar abundances of
  \cite{Asplund+2021}}\label{fig:abu}
\end{figure}

To verify if the two hotter stars in NGC\,6388 (5235 and 7788),
  which show higher than expected iron abundances, are affected by
  diffusion I compare in Fig.\,\ref{fig:abu} the abundances of the
stars in NGC\,6388 to those found for stars affected by diffusion in
NGC\,6397 and NGC\,6752. While the iron and chromium abundances
overlap closely, the abundances of silicon, manganese, and yttrium
differ substantially, ruling out diffusion as the explanation of the
higher iron abundances found for the stars 5235 and (to a lesser
extent) 7788 in NGC\,6388.

\begin{figure}[!h]
\includegraphics[height=\columnwidth,angle=270]{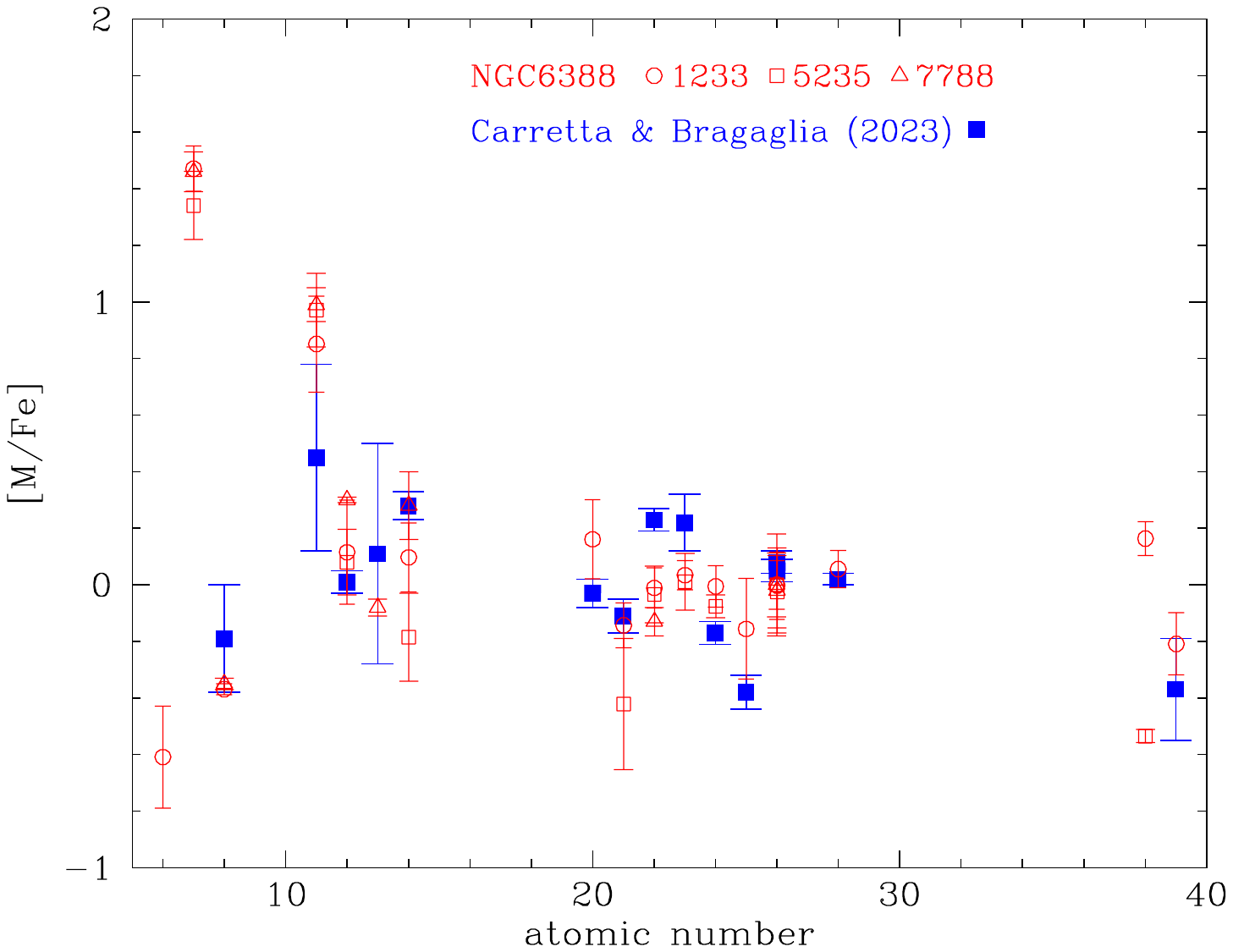}
\caption[]{Abundances of stars in NGC\,6388 relative to iron from this
  paper (red open symbols) and from \cite{Carretta+2023} (blue filled squares, adjusted for the
  different solar abundances used). The error bars for the data from \cite{Carretta+2023} refer to sample errors.}\label{fig:abu_CB23}
\end{figure}

Figure\,\ref{fig:abu_CB23} compares for NGC\,6388 the abundances from my
analysis to those determined for red giant stars in this cluster by
\citet{Carretta+2023}\footnote{I use the solar abundances from
\cite{Asplund+2021}, which differ slightly from those used by
\cite{Carretta+2023}. When comparing my abundances to those of
\cite{Carretta+2023} I adjust for those differences.}.
For most of the elements, the results agree
well, with the blue HB stars often showing abundances at the edges
of the distribution found by \citet{Carretta+2023}. The sodium
and oxygen abundances of the HB stars put them at the extreme end of the `extreme'
population in Fig.\,1 of \citet{Carretta+2023}. In addition to the elements
covered by \cite{Carretta+2023}, the HB stars show low carbon and high
nitrogen abundances.
Magnesium and aluminium, on the other hand, do not show extreme abundances.

In summary, the observed abundances of carbon, oxygen, nitrogen, and sodium are consistent with the abundance patterns expected for products of hot hydrogen burning, which might produce also helium enrichment. On the other hand, magnesium and aluminium do not show significant differences from the heavy element abundances, so  this stronger requirement for helium enrichment has not been satisfied.

\subsection{Diffusion: Abundance changes with temperature}\label{ssec:rel_abu_diffusion}
\begin{figure}[!h]
\includegraphics[width=0.9\columnwidth,angle=0]{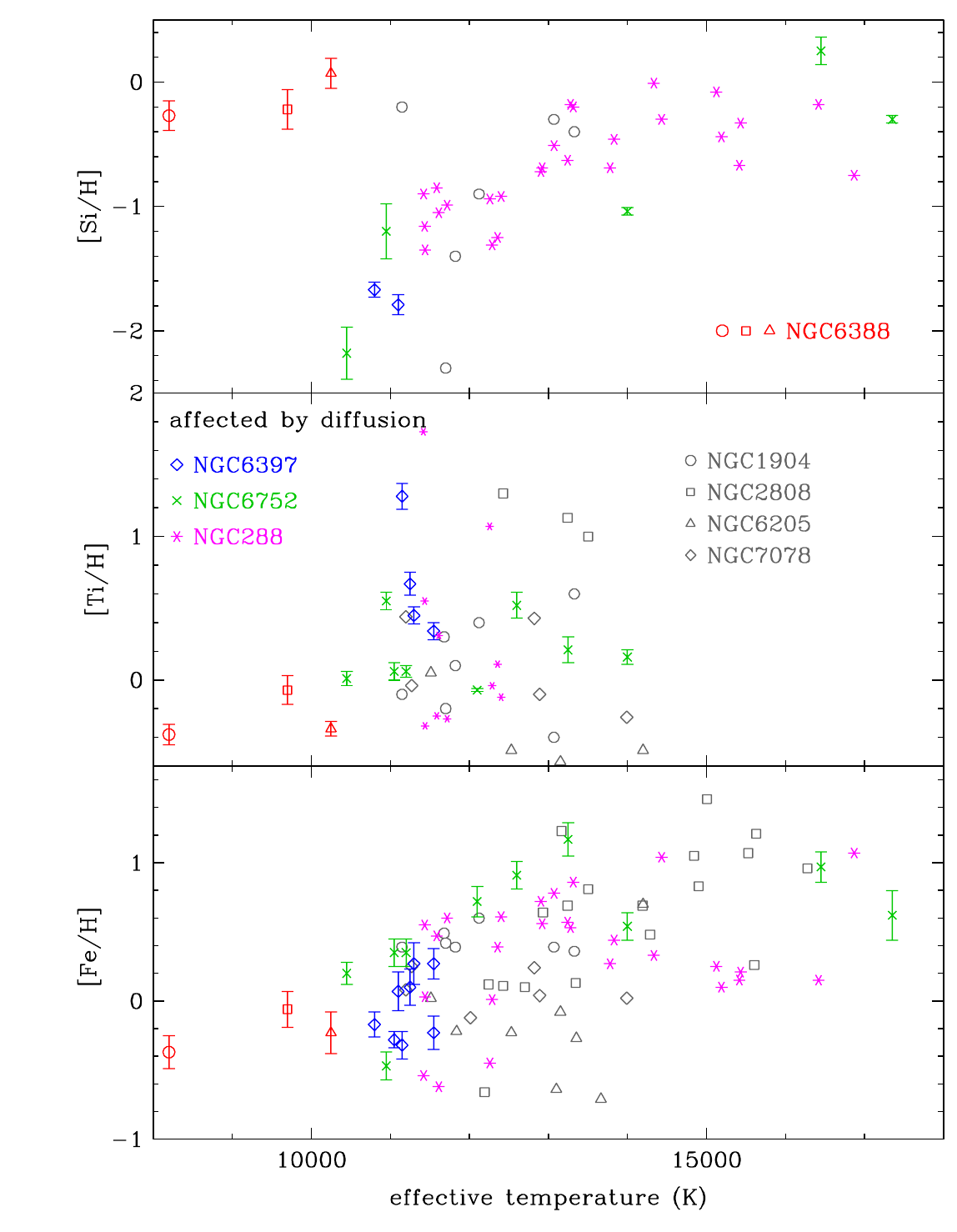}
\caption[]{  
Abundances of silicon, titanium, and iron (relative
    to the solar abundances of \cite{Asplund+2021}) vs effective
    temperature for stars in NGC\,6388 (red symbols) and stars
  affected by diffusion in NGC\,288 (magenta asterisks,
  \citealt{Moehler+2014}), NGC\,6397 (blue lozenges), and
    NGC\,6752 (green crosses). The grey symbols mark abundances
    for HB stars in NGC\,1904 \citep{Fabbian+2005}, NGC\,2808
    \citep{Pace+2006}, NGC\,6205, and NGC\,7078
    \citep{behr03}.}\label{fig:abu_iron}
\end{figure}

Figure\,\ref{fig:abu_iron} shows the abundances of silicon,
  titanium, and iron for the blue HB stars affected by diffusion in
  NGC\,6397, NGC\,6752, NGC\,288 \citep{Moehler+2014}, NGC\,1904
  \citep{Fabbian+2005}, NGC\,2808 \citep{Pace+2006}, NGC\,6205, and
  NGC\,7078 \citep{behr03}.  For comparison, I also included the
  abundances of the blue HB stars in NGC\,6388 that are unaffected by
  diffusion.

One obvious common feature are the large star-to-star variations in
abundance around the Grundahl jump, namely, at the onset of radiative
levitation. The data from \citet[open grey
  circles]{Fabbian+2005} tend to follow the data from NGC\,288
\citep{Moehler+2014} and NGC\,6752. The data from NGC\,6205 (open grey
triangles) and NGC\,7078 (open grey lozenges), both from
\citet{behr03} instead tend to lie at the lower part of the
distribution, while those from NGC\,2808 \citep[open
  squares]{Pace+2006} occupy the upper parts of the abundance
distribution. For these three clusters, there are no abundances for
silicon.  NGC\,2808 is the most metal-rich cluster for
which abundances for hot blue HB stars are available ([M/H] = $-$1.14,
\citealt{Harris96}), while NGC\,1905 ([M/H] = $-$1.60,
\citealt{Harris96}), NGC\,6205([M/H] = $-$1.53, \citealt{Harris96}),
and NGC\,7078 ([M/H] = $-$2.37, \citealt{Harris96}) have similar
metallicities to NGC\,6752 and NGC\,6397.
It is unclear how much of the differences may be due to different
data and analysis methods.

The different elements shown
  in Fig.\,\ref{fig:abu_iron} behave differently with effective
  temperature: titanium shows mainly large scatter
  without clear trends, possibly due to the fact that it is found
  only in a subsample of all stars. Silicon shows increasing
  abundances with higher effective temperatures, which turns into a
  flat distribution around [Si/H] of about $-$0.4 (albeit with large
  scatter) above temperatures of about 13\,000\,K. Because the silicon data are available only for globular clusters with moderately low metallicities ([M/H] = $-$1.32\ldots$-$1.60), it is not clear if the cluster metallicity has any effect.

  For iron the scatter at temperatures below about 14\,000\,K is also
  large, but caused mostly by the low values for NGC\,6205 (open grey
  triangles). Ignoring those, the abundances increase for effective
  temperatures of up to 14\,000\,K and then split into a bimodal
  distribution, with part of the stars following a trend of decreasing
  iron abundance with increasing temperature and the other part
  showing a roughly constant iron abundance of about [Fe/H] = +1 for
  the same temperature range. It is currently unclear what creates
  these two branches.

\subsection{Notes to other observers looking for helium enrichment}\label{ssec:notes}
 The high-resolution spectra of cool blue HB stars in metal-rich
  globular clusters are very well suited to determine metal abundances
  and atmospheric parameters.  For the measurement of helium
  abundances, on the other hand, a high signal-to-noise ratio and a smooth
  continuum are more important than having high resolution. This is because
  the helium lines are inherently broadened and therefore  wider
  and shallower than metal lines.

For cool blue HB stars in metal-poor globular clusters, the situation
is quite different because the predicted helium enrichment is
generally small and they show very few metal lines. If atmospheric
parameters can be determined from other data, medium-resolution spectra with high signal-to-noise are better suited
to measure helium abundances for these stars as well, rather
than high-resolution spectra with lower
signal-to-noise ratios.  

Overall, the study of the effects of radiative levitation in the moderately hot
HB stars (effective temperatures between about 11\.000\,K and
17\,000\,K) benefits from high resolution thanks to
the many metal lines available, independently of the original metallicity.

\section{Conclusions}
Blue HB stars in metal-rich globular clusters can provide important
constraints on pollution scenarios because they have a high
likelihood of being affected by pollution.  The results presented here
for NGC\,6388 may serve as additional constraints when it comes to
understanding the status of this exceptional globular
cluster. They point towards the presence of products of hot
hydrogen-burning in the blue HB stars, which could indicate
accompanying helium enrichment; however, the situation is not entirely clear-cut. Additional data (see
Sect.\,\ref{ssec:notes}) could help  improve the constraints.

The abundances found for silicon, titanium, and iron in moderately hot stars show different trends with effective temperatures, which may indicate
the effect of parameters beyond rotation and effective
temperature. The question of whether the helium abundance plays a role could be
studied in such clusters as NGC\,6388 and NGC\,6441, for which higher
effective temperatures are predicted for the Grundahl jump.

\section{Data availability}
The flux calibrated spectra, the measured equivalent widths, and the line parameters used to perform the abundance analyses are only available in electronic form at the Strasbourg astronomical Data Centre CDS via anonymous ftp to cdsarc.u-strasbg.fr (130.79.128.5) or via http://cdsweb.u-strasbg.fr/cgi-bin/qcat?J/A+A/.
  \begin{acknowledgements}
  The work reported here could not have succeeded without the support of many people. I want to thank them here in the sequence of the time at which they helped me: J. Pritchard for his help in
  understanding some of the problems I faced with the red arm data, A. Mucciarelli for his patience and support with GALA,  S. Villanova for providing me with some testing data of blue HB stars in NGC\,6752, E. Pancino for her support in using Daospec and understanding its behaviour, A. Bragaglia and E. Carretta for providing advice about abundances in NGC\,6388 and encouragement towards a publication, and, last but by now means least, W.V.D Dixon, whose referee report improved the content and the readability of this paper immensely.

  This investigation also could not have been performed without the
  many services listed below.  This research has made use of the
  NASA's Astrophysics Data System Bibliographic Services. This work has made
  use of the VALD database, operated at Uppsala University, the
  Institute of Astronomy RAS in Moscow, and the University of Vienna
  \citep{VALD3}, and of data from the European Space Agency (ESA)
  mission {\it Gaia} (\url{https://www.cosmos.esa.int/gaia}),
  processed by the {\it Gaia} Data Processing and Analysis Consortium
  (DPAC,
  \url{https://www.cosmos.esa.int/web/gaia/dpac/consortium}). Funding
  for the DPAC has been provided by national institutions, in
  particular the institutions participating in the {\it Gaia}
  Multilateral Agreement, and of the VizieR catalogue access tool,
  CDS, Strasbourg, France. The original description of the VizieR
  service was published in \cite{VizieR}.  Some of the data presented
  in this paper were obtained from the Mikulski Archive for Space
  Telescopes (MAST). STScI is operated by the Association of
  Universities for Research in Astronomy, Inc., under NASA contract
  NAS5-26555.
\end{acknowledgements}
\bibliographystyle{aa}
\bibliography{NGC6388_VALD}
\begin{appendix}
  \section{Observing information}\label{sec:app_obs_info}
  This section presents the colour-magnitude diagrams of the blue
  horizontal branches in the three globular clusters, NGC\,6388,
  NGC\,6397, and NGC\,6752, along with the coordinates, observing times and
  conditions together with the optical photometric information for the
  stars discussed in this publication.\nopagebreak

  \subsection{Colour-magnitude diagrams}\label{app:cmd}
  Here, the colour-magnitude diagrams are shown, based on which the targets have been selected.
  \begin{figure}[!h]
\includegraphics[height=\columnwidth,angle=270]{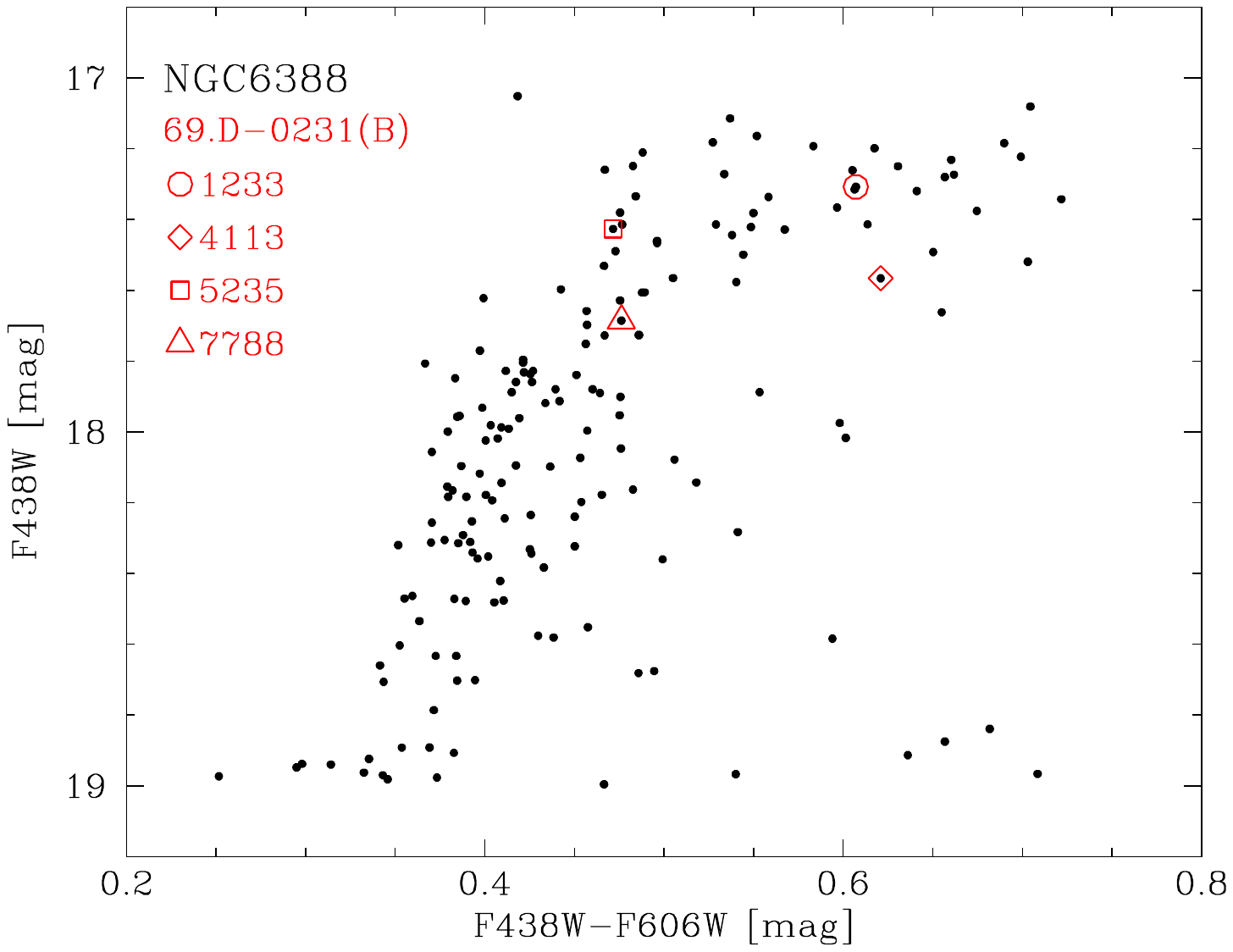}
\caption[]{Colour-magnitude diagram for the globular cluster NGC\,6388
  using data from the HST UV Globular Cluster Survey (HUGS,
  \citealt{Piotto+HUGS},
  \citealt{Nardiello+HUGS}). The red symbols mark the targets of run 69.D-0231(B)}\label{fig:NGC6388_cmd}
\end{figure}

  \begin{figure}[!h]
\includegraphics[height=\columnwidth,angle=270]{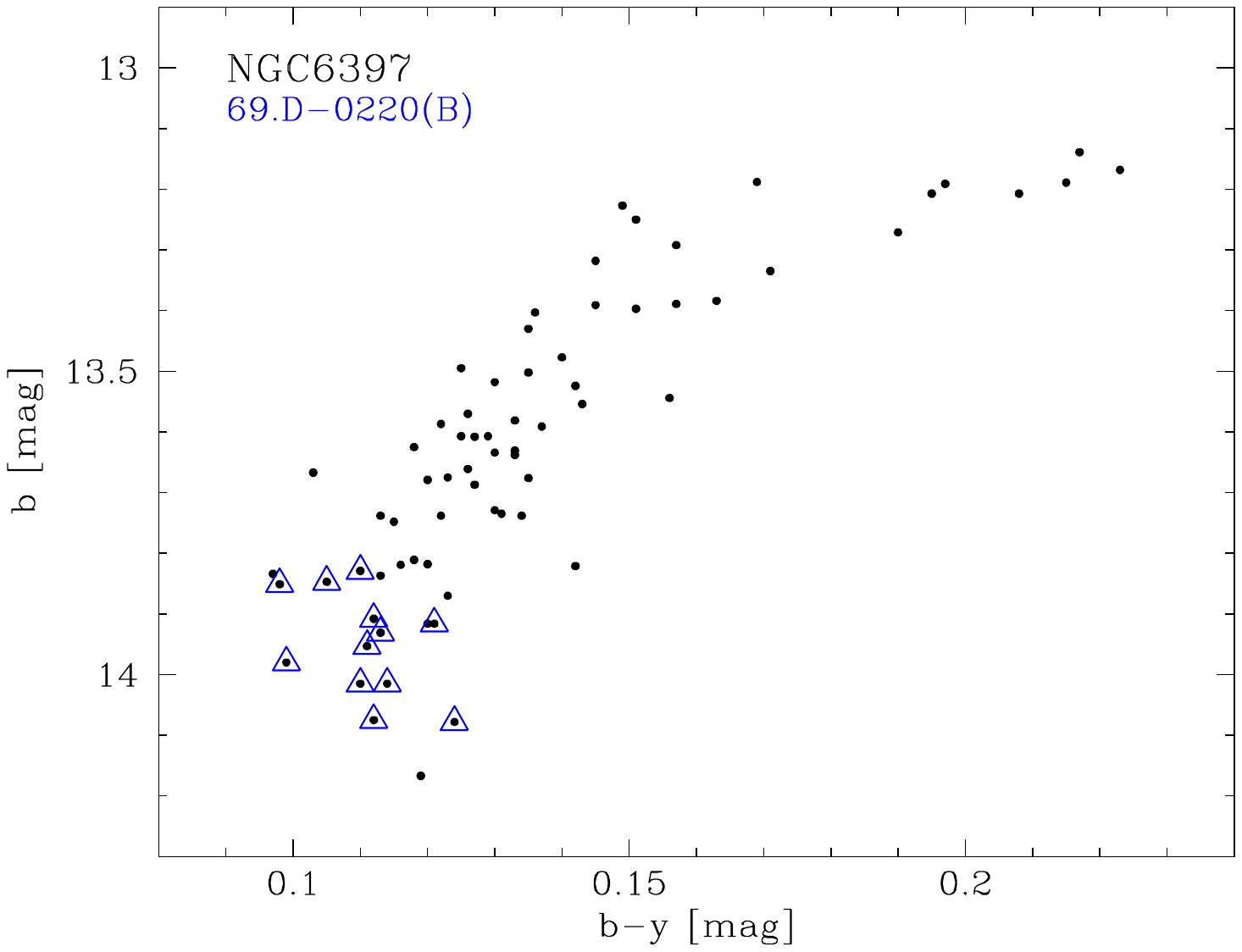}
\caption[]{Colour-magnitude diagram for the globular cluster NGC\,6397
  using Str\"omgren photometry from \citet{Twarog2000}. The blue triangles mark the targets of run 69.D-0220(B).}\label{fig:NGC6397_cmd}
\end{figure}

  \begin{figure}[!h]
\includegraphics[height=\columnwidth,angle=270]{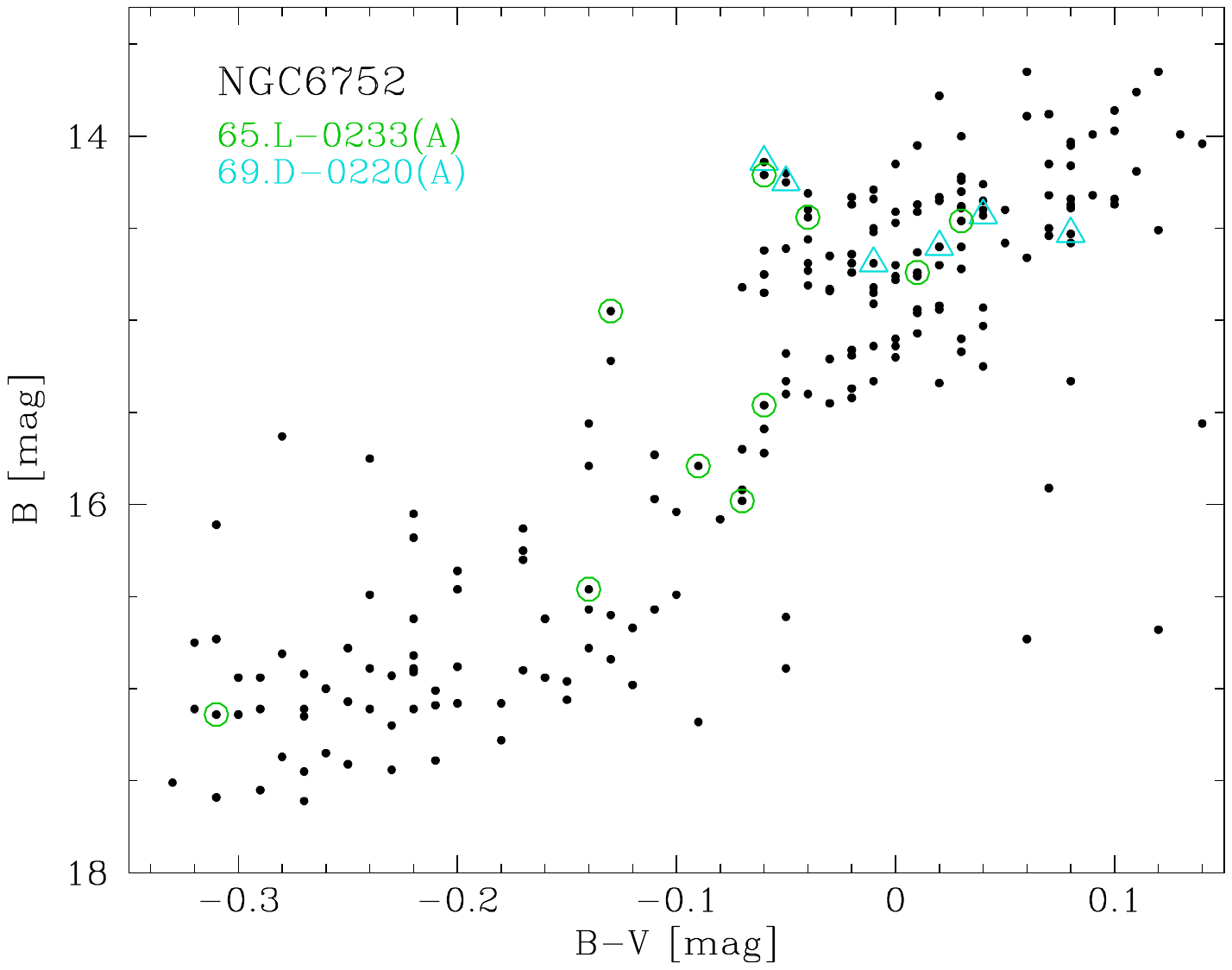}
\caption[]{Colour-magnitude diagram for the globular cluster NGC\,6752
  using Johnson photometry from \citet{Buonanno+86}. The green circles mark the targets of the run 65.L-0233 and the cyan triangles mark the targets of the run 69.D-0220(A).}\label{fig:NGC6752_cmd}
\end{figure}
\onecolumn
  \subsection{Target and observing information}\label{app:targets}\nopagebreak
  Here, the coordinates and photometric information for the targets, together with the details of their observations, are listed.
\begin{table*}[!h]
  \caption[]{Target coordinates \citep{Gaiaproject,GaiaEDR3} and
    photometric data (from \citealt{Piotto+97}) for NGC\,6388 together
    with observing information. The seeing is the average value
    obtained by the Differential Image Motion Monitor (DIMM),
    corrected to zenith.}\label{tab:targobs}
\begin{tabular}{lllrrllll}
\hline
\hline
star & $\alpha_{2000}$ & $\delta_{2000}$ & $V$ & $B-V$ 
& start & exposure time & \multicolumn{2}{c}{average}\\
     &                 &               &      &
     &       &               & airmass & seeing \\
     &  [h:m:s]        & [$^\circ$:\arcmin:\arcsec]  &[mag] &[mag] 
     &  [UT]        & [sec]  &         & [''] \\
\hline
 1233 & 17:36:24.575 & $-$44:43:15.28 & 16.867$\pm$0.029 & +0.428$\pm$0.038
 & 2002-06-15T05:56 & 3000 &  1.14 &  1.33 \\
 & & & & & 2002-06-15T06:47 & 2500 &  1.23 &  1.76 \\
 & & & & & 2002-06-16T00:43 & 3000 &  1.49 &  0.54 \\
 & & & & & 2002-06-16T03:49 & 3000 &  1.07 &  1.14 \\
 & & & & & 2002-07-10T00:36 & 3000 &  1.19 &  0.70 \\
 & & & & & 2002-07-10T03:47 & 3000 &  1.10 &  0.97 \\
 & & & & & 2002-08-03T02:20 & 3000 &  1.11 &  1.21 \\
 & & & & & 2002-08-03T03:12 & 3000 &  1.19 &  1.40 \\
 4113 & 17:36:22.879 & $-$44:44:24.68 & 16.776$\pm$0.026 & +0.545$\pm$0.037
 & 2002-06-15T00:52 & 3000 &  1.46 &  1.01 \\
 & & & & & 2002-06-15T01:43 & 3000 &  1.27 &  1.11 \\
 & & & & & 2002-06-15T04:05 & 3000 &  1.07 &  1.11 \\
 & & & & & 2002-06-15T04:57 & 3000 &  1.08 &  1.31 \\
 5235 & 17:36:21.289 & $-$44:45:13.98 & 17.113$\pm$0.019 & +0.341$\pm$0.024
 & 2002-04-07T08:44 & 2700 &  1.07 &  0.62 \\
 & & & & & 2002-05-06T08:07 & 2700 &  1.10 &  0.67 \\
 & & & & & 2002-05-06T08:53 & 2700 &  1.17 &  0.60 \\
 & & & & & 2002-05-09T07:27 & 2700 &  1.08 &  0.65 \\
 & & & & & 2002-05-09T09:19 & 2700 &  1.25 &  0.76 \\
 7788 & 17:36:21.846 & $-$44:42:55.64 & 17.410$\pm$0.023 & +0.360$\pm$0.026
 & 2002-05-18T06:02 & 2820 &  1.07 &  1.05 \\
 & & & & & 2002-05-18T06:50 & 2820 &  1.08 &  1.17 \\
 & & & & & 2002-06-02T03:30 & 2820 &  1.15 &  0.85 \\
 & & & & & 2002-06-10T02:36 & 2820 &  1.19 &  0.76 \\
 & & & & & 2002-06-12T03:31 & 2820 &  1.10 &  1.01 \\
 & & & & & 2002-06-13T02:09 & 2820 &  1.22 &  1.17 \\ 
 & & & & & 2002-06-14T03:51 & 2820 &  1.08 &  0.80 \\ 
\hline
\end{tabular}
\end{table*}

\begin{table*}[!h]
  \caption[]{Target coordinates \citep{Gaiaproject,GaiaEDR3} and
    photometric data for NGC\,6397 \citep{Twarog2000}. The seeing is
    the average value obtained by the DIMM), corrected to
    zenith.\label{tab:targobs6397}}
\begin{tabular}{lllrrllll}
\hline
\hline
star & $\alpha_{2000}$ & $\delta_{2000}$ & $V$ & $b-y$ & start & exposure & \multicolumn{2}{c}{average}\\
     &                 &               &      &       &       &  time   & airmass & seeing \\
     &  [h:m:s]        & [$^\circ$:\arcmin:\arcsec]  &[mag] &[mag] &  [UT]        & [sec]  &         & [''] \\
\hline
T160 & 17:40:55.005 & $-$53:40:07.19& 13.719$\pm$0.003 & $+$0.110$\pm$0.004 & 2002-04-25T07:20:58 & 1500 & 1.15 & 1.23 \\
T164 & 17:40:43.912 & $-$53:31:00.85& 13.742$\pm$0.003 & $+$0.105$\pm$0.004 & 2002-04-25T07:50:55 & 1680 & 1.14 & 1.10 \\
T166 & 17:40:34.642 & $-$53:28:49.36& 13.753$\pm$0.004 & $+$0.098$\pm$0.004 & 2002-04-25T06:14:16 & 1680 & 1.21 & 1.37 \\
T170 & 17:41:12.138 & $-$53:38:29.34& 13.795$\pm$0.002 & $+$0.121$\pm$0.003 & 2002-04-28T05:47:15 & 1800 & 1.23 & 1.88 \\
     &             &              &                  &                    & 2002-05-10T07:27:01 & 1800 & 1.16 & 1.43 \\
T172 & 17:40:02.178 & $-$53:39:43.87& 13.796$\pm$0.002 & $+$0.112$\pm$0.003 & 2002-04-25T03:50:50 & 1800 & 1.65 & ---  \\
T174 & 17:40:39.477 & $-$53:36:44.80& 13.818$\pm$0.003 & $+$0.113$\pm$0.002 & 2002-04-25T06:46:31 & 1800 & 1.18 & 1.24 \\
T179 & 17:40:55.468 & $-$53:37:12.11& 13.842$\pm$0.003 & $+$0.111$\pm$0.004 & 2002-04-25T09:08:02 & 1800 & 1.19 & 1.21 \\
T183 & 17:40:20.065 & $-$53:41:43.27& 13.881$\pm$0.002 & $+$0.099$\pm$0.004 & 2002-04-25T05:02:40 & 1860 & 1.37 & ---  \\
T185 & 17:40:26.707 & $-$53:37:52.69& 13.901$\pm$0.003 & $+$0.114$\pm$0.004 & 2002-04-25T05:39:04 & 1860 & 1.26 & 1.30 \\
T186 & 17:40:19.822 & $-$53:36:30.38& 13.905$\pm$0.004 & $+$0.110$\pm$0.004 & 2002-04-25T04:27:10 & 1860 & 1.49 & --- \\
T191 & 17:40:56.029 & $-$53:35:07.66& 13.954$\pm$0.002 & $+$0.124$\pm$0.003 & 2002-04-14T08:49:20 & 2100 & 1.14 & 0.94 \\
T193 & 17:40:57.467 & $-$53:35:33.07& 13.963$\pm$0.002 & $+$0.112$\pm$0.002 & 2002-04-25T08:23:17 & 2100 & 1.16 & 1.10  \\
\end{tabular}
\end{table*}

\begin{table*}[!h]
  \caption[]{Target coordinates \citep{Gaiaproject,GaiaEDR3} and
    photometric data for NGC\,6752
    \citep{Buonanno+86}) together with observing information. The
    seeing is the average value obtained by the DIMM, corrected to
    zenith.\label{tab:targobs6752}}
\begin{tabular}{lllrrllll}
\hline
\hline
star & $\alpha_{2000}$ & $\delta_{2000}$ & $V$ & $B-V$ & start & exposure & \multicolumn{2}{c}{average}\\
     &                 &               &      &       &       &  time   & airmass & seeing \\
     &  [h:m:s]        & [$^\circ$:\arcmin:\arcsec]  &[mag] &[mag] &  [UT]        & [sec]  &         & [''] \\
  \hline
\multicolumn{9}{c}{65.L-0233}\\
  \hline
B491 & 19:11:36.751 & $-$60:03:12.84 & 17.45 & $-$0.31 & 2000-05-22T05:01:59 & 7200 & 1.31 & 0.45\\
     &              &                &       &         & 2000-05-22T07:14:32 & 7080 & 1.24 & 0.39\\
     &              &                &       &         & 2000-05-23T05:34:12 & 3540 & 1.30 & 1.18\\
     &              &                &       &         & 2000-05-23T08:33:21 & 3540 & 1.27 & 0.93\\
B1509& 19:11:14.277 & $-$59:54:48.42 & 15.52 & $-$0.06 &  2000-04-24T07:32:24 & 4320 & 1.28 & 1.09 \\
     &              &                &       &         & 2000-05-18T05:12:17 & 4320 & 1.35 & 1.26 \\
     &              &                &       &         & 2000-05-21T07:59:08 & 4320 & 1.24 & 0.43 \\
B2099& 19:11:02.982 & $-$59:55:55.10 & 15.88 & $-$0.09 & 2000-04-25T06:21:04 & 6120 & 1.38 & 1.68 \\
     &              &                &       &         & 2000-05-20T04:40:49 & 6120 & 1.38 & 0.54 \\
B2454& 19:10:56.455 & $-$60:05:26.47 & 14.27 & $-$0.06 & 2000-05-19T04:43:19 & 1440 & 1.46 & 0.74 \\
B2698& 19:10:50.666 & $-$60:02:39.17 & 15.08 & $-$0.13 & 2000-04-24T08:56:38 & 3240 & 1.23 & 1.33\\
     &              &                &       &         & 2000-05-19T05:20:09 & 3240 & 1.35 & 0.65 \\
B2735& 19:10:49.913 & $-$60:04:08.14 & 14.43 & $+$0.03 & 2000-05-19T06:21:52 & 1800 & 1.27 & 0.65\\
B3699& 19:10:30.056 & $-$59:58:24.39 & 16.05 & $-$0.07 & 2000-05-20T06:36:18 & 7200 & 1.24 & 0.62 \\
B4172& 19:10:20.205 & $-$59:57:40.18 & 14.48 & $-$0.04 & 2000-05-19T07:00:19 & 1800 & 1.24 & 0.55\\
B4548& 19:10:12.035 & $-$59:51:55.84 & 16.60 & $-$0.14 & 2000-04-25T08:12:45 & 2700 & 1.25 & 1.85\\
     &              &                &       &         & 2000-04-25T09:05:50 & 2700 & 1.23 & 1.94\\
     &              &                &       &         & 2000-05-21T04:34:15 & 6000 & 1.38 & 0.48\\
     &              &                &       &         & 2000-05-21T06:23:08 & 5400 & 1.24 & 0.42\\
B4598& 19:10:10.912 & $-$59:59:41.60 & 14.73 & $+$0.01 & 2000-05-19T07:35:43 & 2280 & 1.23 & 0.64\\
\hline
\multicolumn{9}{c}{69.D-0220(A)}\\
\hline
B652  & 19:11:31.496 & $-$59:57:52.54 & 14.70 & $-$0.01 & 2002-06-15T08:37:57 & 2130 & 1.41 & 1.60 \\
      &              &                &       & & 2002-06-15T09:20:01 & 2130 & 1.53 & 1.87 \\
B944  & 19:11:25.779 & $-$59:56:17.12 & 14.58 & $+$0.02 & 2002-06-15T07:30:40 & 1770 & 1.28 & 1.54 \\
      &              &                &       & & 2002-06-15T08:07:00 & 1770 & 1.33 & 1.40 \\
B2151 & 19:11:01.963 & $-$59:55:30.96 & 14.45 & $+$0.08 & 2002-06-13T05:32:28 & 2280 & 1.24 & 1.30  \\
      &              &                &       & & 2002-07-31T02:25:50 & 2280 & 1.23 & 0.83 \\
B2206 & 19:11:00.928 & $-$59:57:42.29 & 14.30 & $-$0.05 & 2002-06-13T04:58:35 & 1620 & 1.26 & 1.20 \\
      &              &                &       & & 2002-07-31T03:53:09 & 1620 & 1.25 & 0.87 \\
B2281 & 19:10:59.395 & $-$59:57:06.05 & 14.39 & $+$0.04 & 2002-06-13T03:37:52 & 1980 & 1.37 & 1.28 \\
      &              &                &       & & 2002-07-31T03:17:23 & 1980 & 1.23 & 0.83  \\
B2649 & 19:10:51.610 & $-$60:02:26.05 & 14.35 & $-$0.02 & 2002-06-13T04:24:30 & 1740 & 1.30 & 1.38  \\
      &              &                &       & & 2002-07-31T04:31:14 & 1740 & 1.28 & 0.81 \\
B3243 & 19:10:39.075 & $-$60:03:12.07 & 14.20 & $-$0.06 & 2002-06-13T06:16:41 & 1140 & 1.23 & 1.10 \\
      &              &                &       & & 2002-07-31T05:07:51 & 1140 & 1.33 & 0.76 \\
\hline
\end{tabular}
\end{table*}
\clearpage
\twocolumn
\section{Observing issues with the data for NGC\,6388}\label{app_ngc6388_obs}
Here I describe the problems I encountered with some of the observations obtained for NGC\,6388 and how they were solved.
\begin{figure}[!h]
\includegraphics[height=\columnwidth,angle=270]{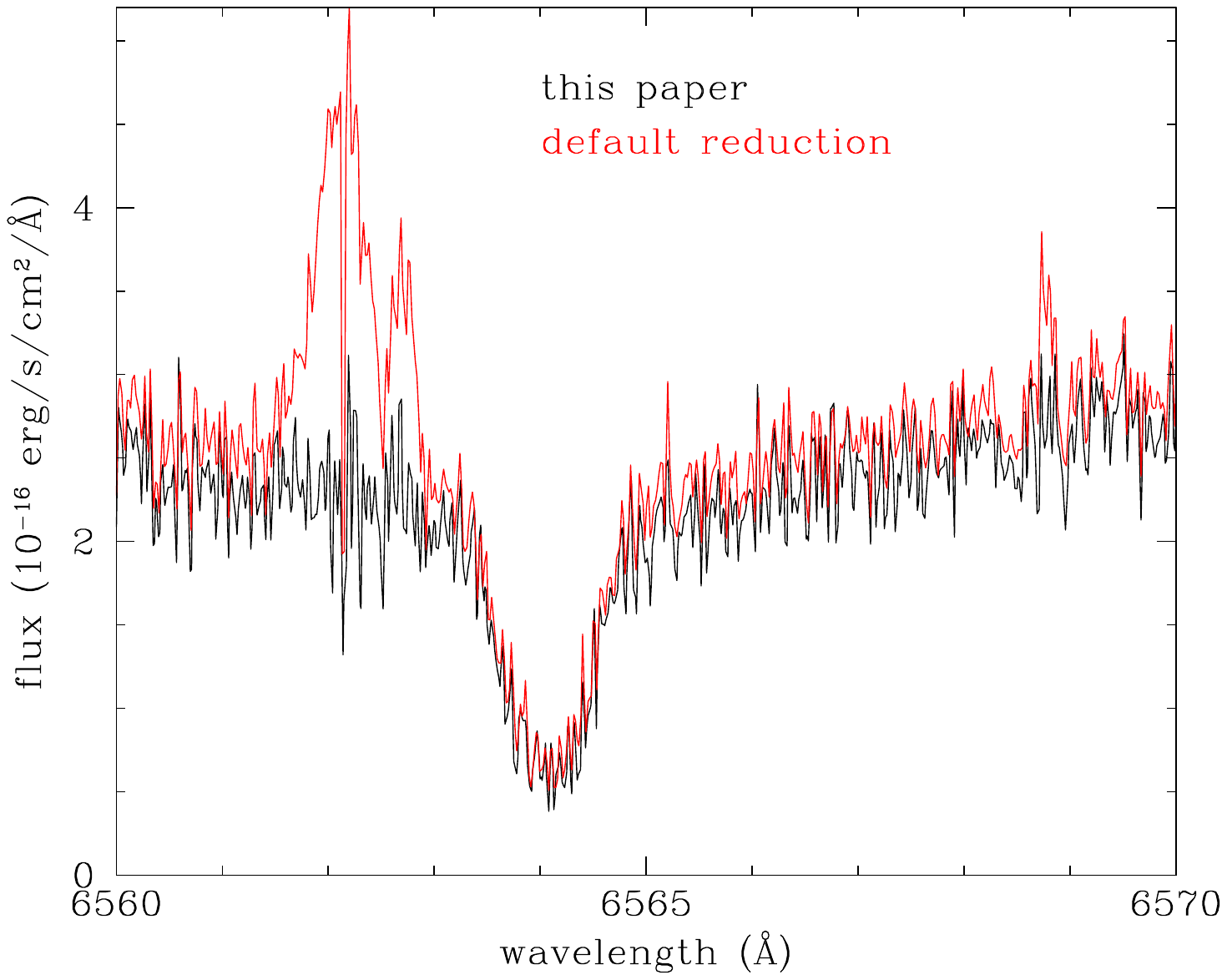}
\includegraphics[height=\columnwidth,angle=270]{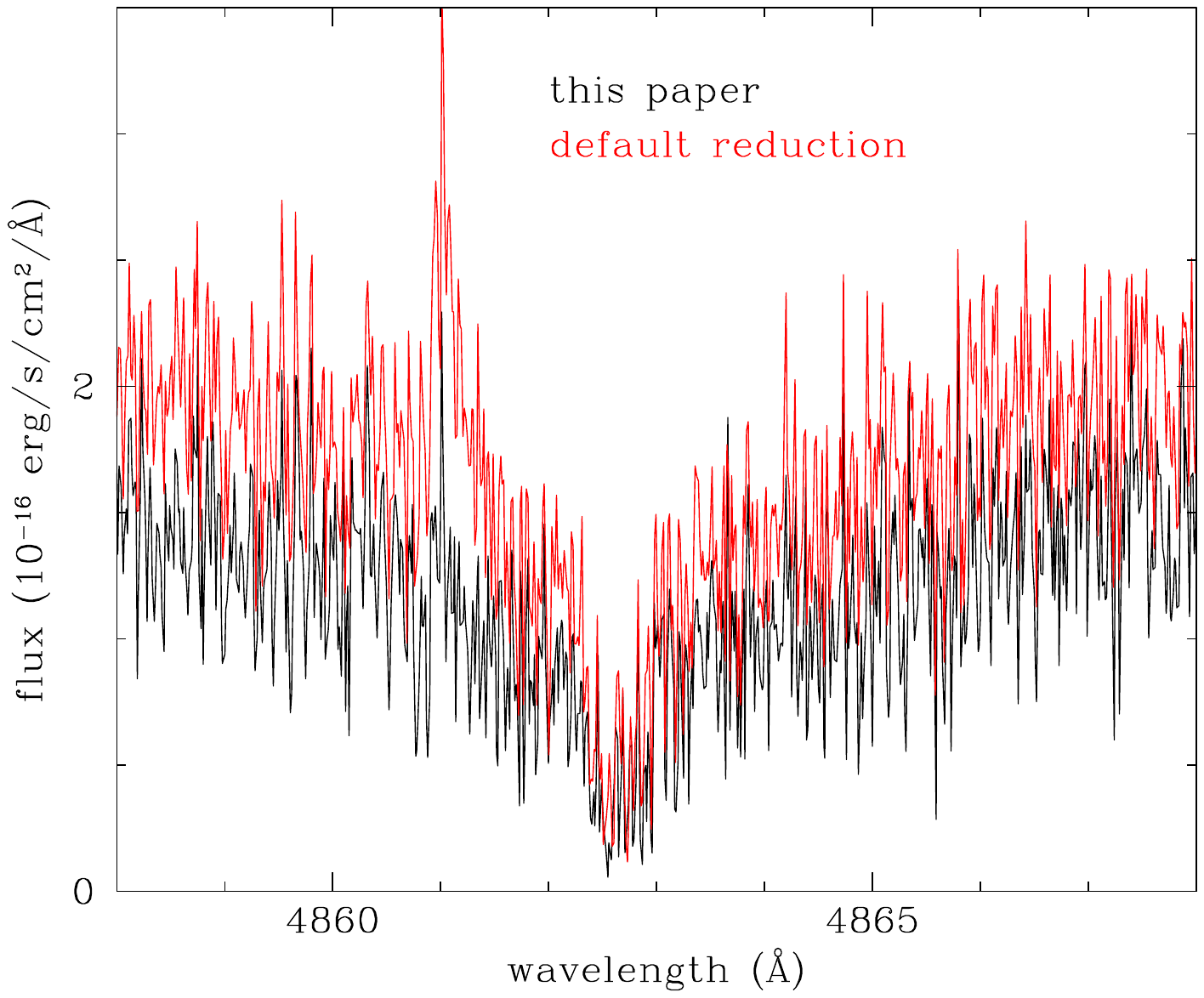}
\caption[]{Top: H$\alpha$ region of the second observation of
  5235 from 5 May 2002, reduced as described here (black) and with
  default reduction parameters (red). One can clearly see the spurious emission
  on the blue side of the stellar H$\alpha$ line
  core. Bottom: H$\beta$ region of the observation of
  7788 from 12 June 2002, reduced as described here (black) and with
  default reduction parameters (red). One can clearly see the spurious emission
  on the blue side of the stellar H$\beta$ line
  core.}\label{fig:emis}
\end{figure}

\begin{description}
\item [{\bf 1233}] The data observed for this star on the night
  starting on 15 June 2002 show a shift of about 100\,pixels (comparable to
  the distance between orders) with respect to the daytime
  calibrations. Fortunately nighttime calibrations had been observed
  which could be used to process the data, which showed a shift in
  wavelength of some 50\,\AA\ when reduced with daytime calibrations.

\item [{\bf 5235}] The spectra from 5 May 2002 (start of the night)
  show blue-shifted emission at H$\alpha$ (and to a lesser extent at
  H$\beta$), which is surprising as cool blue HB stars are not known
  for emission lines. A closer inspection revealed that the orders of
  the spectrum had shifted between the science observations and the
  corresponding daytime calibrations, which caused an imperfect sky
  subtraction and retained background H$\alpha$ emission in the
  extracted spectrum. By using the science spectra to identify the
  order centre and reducing the slit length for the spectrum
  extraction to 20\,pixels this problem could be solved (see
  Fig.\,\ref{fig:emis}, top).
\item [{\bf 7788}] Despite the attempts to select isolated targets a
  faint additional spectrum appears in the slit for the red arm. To
  avoid contamination by this additional source this the slit length
  was again reduced to 20\,pixels. In addition the data observed for
  this star on 11, 12, and 13 June 2002 (start of the night) suffer
  again from order shifts between science data and daytime
  calibrations, which cause spurious emission in the line cores of
  H$\alpha$ and H$\beta$. This problem could be solved for the data
  from June 11 and 12 by using the science spectra to identify the
  order centre (see Fig.\,\ref{fig:emis}, bottom) . The shift of the
  science data from June 13 was too large to recover the red part of
  the red arm.
\end{description}
\section{Atmospheric parameters derived from line profile fitting}\label{sec:param_fit}

Here, I provide the results of the line profile fitting process
described in Sect.\,\ref{sec:params}.
\begin{table}[ht]
  \caption[]{Atmospheric parameters for the stars in NGC\,6397 (T$<$number$>$) and
    NGC\,6752 (B$<$number$>$) from fitting the hydrogen and (if available) helium line
    profiles for stars without evidence of diffusion. The values in
    italics are taken from
    \citet[ Table 1]{mosw00}.}\label{tab:metal_poor_param}
  \begin{tabular}{llrr}
    \hline
    \hline
    star & $T_{eff}$ & $\log g$ & $\log \frac{n_{He}}{n_{H}}$\\
    & [K] & [cm sec$^{-2}$] & \\
    \hline
    \multicolumn{4}{c}{stars unaffected by diffusion}\\
    \hline
    \multicolumn{4}{c}{65.L-0233(A) (NGC\,6752)}\\
    \hline
B2454 &  9340$\pm$80 & 3.12$\pm$0.04 &  \\
B2735 & 10260$\pm$70 & 3.36$\pm$0.02 & \\
     & {\it 11100$\pm$260} & {\it 3.78$\pm$0.12} & {\it $-$1.14$\pm$0.36}\\ [1mm]  
    \hline
    \multicolumn{4}{c}{69.D-0220(A) (NGC\,6752)}\\
    \hline
 B944  & 10400$\pm$50 & 3.50$\pm$0.02&  \\
 & {\it 11100$\pm$230} & {\it 3.70$\pm$0.10} & {\it $-$0.84$\pm$0.31 }\\ [1mm]  
 B2281 & 11050$\pm$40 & 3.64$\pm$0.02&  \\
    \hline
    \multicolumn{4}{c}{69.D-0220(B) (NGC\,6397)}\\
    \hline
 T160  & 10080$\pm$50 & 3.45$\pm$0.02&  \\
 T164  &  9950$\pm$40 & 3.43$\pm$0.02&  \\
 T166  & 10000$\pm$60 & 3.43$\pm$0.02&  \\
 T174  & 10440$\pm$40 & 3.57$\pm$0.02&  \\
 \hline
  \end{tabular}
\end{table}

\begin{table*}
  \caption[]{Atmospheric parameters for the stars affected by
    diffusion in NGC\,6397 and NGC\,6752 from fitting the hydrogen and
    (if available) helium line profiles and from the GALA analysis.
    The values in italics are taken from \citet[ Table
      2]{mosw00}}.\label{tab:metal_poor_param_diff}
  \begin{tabular}{l|lrr|lr}
    \hline
    \hline
    star & \multicolumn{3}{c}{line profile fits} & \multicolumn{2}{c}{GALA}\\
    & $T_{eff}$ & $\log g$ & $\log \frac{n_{He}}{n_{H}}$ & $T_{eff}$ & $\log g$ \\
    & [K] & [cm sec$^{-2}$] & & [K] & [cm sec$^{-2}$] \\
    \hline
    \hline
    \multicolumn{6}{c}{65.L-0233(A) (NGC\,6752)}\\
    \hline
  B491  & 25200$\pm$310 & 4.96$\pm$0.04 & $-$2.75$\pm$0.02 & & \\ 
       & {\it 28100$\pm$540} & {\it 5.40$\pm$0.07} & {\it $\le-$3.0} &  & \\ [1mm]  
 B1509 & 15400$\pm$170 & 4.09$\pm$0.04 & $-$2.33$\pm$0.04 & 16450 & 4.55\\
       & {\it 16400$\pm$510} & {\it 4.07$\pm$0.09} & {\it $-$2.15$\pm$0.16} & & \\   [1mm]   
 B2099 & 17400$\pm$210 & 4.44$\pm$0.04 & $-$2.30$\pm$0.04 & 17350 & 4.45\\
       & {\it 18800$\pm$790} & {\it 4.58$\pm$0.10} & {\it $-$2.36$\pm$0.22} & & \\    [1mm]  
 B2698 & 14500$\pm$140 & 4.13$\pm$0.02 & $-$2.49$\pm$0.04 & 14000 & 4.30 \\
       & {\it 14700$\pm$490} & {\it 4.13$\pm$0.10} & {\it $-$2.10$\pm$0.26} & & \\    [1mm]  
 B3699 & 18700$\pm$230 & 4.66$\pm$0.04 & $-$2.18$\pm$0.04 &  & \\
       & {\it 21800$\pm$1050} & {\it 4.60$\pm$0.12} & {\it $-$2.30$\pm$0.10} & &  \\    [1mm]  
 B4172 & 11800$\pm$ 60 & 3.70$\pm$0.02 & $-$2.45$\pm$0.10 & 12600 & 4.10\\
       & {\it 12300$\pm$200} & {\it 3.81$\pm$0.05} & {\it $-$2.49$\pm$0.55} & &  \\    [1mm]  
 B4548 & 20500$\pm$340 & 4.86$\pm$0.04 & $-$1.97$\pm$0.02 & & \\
       & {\it 20700$\pm$1490} & {\it 5.06$\pm$0.19} & {\it $-$2.00$\pm$0.17} & &  \\    [1mm]  
 B4598 & 12170$\pm$ 90 & 3.84$\pm$0.02 & $-$2.91$\pm$0.10 & 13250 & 4.20\\
    \hline
    \multicolumn{6}{c}{69.D-0220(A) (NGC\,6752)}\\
    \hline
  B652 & 11620$\pm$60 & 3.71$\pm$0.02 & $-$1.79$\pm$0.08 & 12100 & 3.90\\
       & {\it 12500$\pm$230} & {\it 3.98$\pm$0.07} & {\it $-$2.19$\pm$0.36} & & \\    [1mm]  
 B2151 & 11090$\pm$60 & 3.72$\pm$0.02 & $-$2.37$\pm$0.10 & 11200 & 3.45\\
 B2206 & 10950$\pm$60 & 3.68$\pm$0.02 & $-$2.25$\pm$0.10 & 11050 & 3.40\\
 B2649 & 11320$\pm$70 & 3.79$\pm$0.02 & $-$2.32$\pm$0.10 & 10950 & 3.35\\
 B3243 & 10630$\pm$60 & 3.59$\pm$0.04 & $-$2.16$\pm$0.12 & 10450 & 3.25\\
    \hline
    \multicolumn{6}{c}{69.D-0220(B) (NGC\,6397)}\\
    \hline
 T170  & 11470$\pm$90 & 3.80$\pm$0.04 & $-$1.43$\pm$0.12 & 11550 & 3.80 \\
 T172  & 10700$\pm$80 & 3.65$\pm$0.04 & $-$2.67$\pm$0.16 & 11100 & 3.75\\
 T179  & 10910$\pm$80 & 3.74$\pm$0.04 & $-$2.57$\pm$0.14 & 10800 & 3.40\\
 T183  & 11320$\pm$60 & 3.72$\pm$0.02 & $-$1.95$\pm$0.08 & 11300 & 3.30 \\
 T185  & 11090$\pm$70 & 3.77$\pm$0.02 & $-$2.28$\pm$0.12 & 11050 & 3.50\\ 
 T186  & 11090$\pm$70 & 3.70$\pm$0.02 & $-$2.08$\pm$0.12 & 11250 & 3.60\\
 T191  & 11480$\pm$80 & 3.79$\pm$0.02 & $-$2.15$\pm$0.12 & 11150 & 3.60\\
 T193  & 11310$\pm$50 & 3.75$\pm$0.02 & $-$2.20$\pm$0.08 & 11550 & 3.60\\
 \hline
  \end{tabular}
\end{table*}
\clearpage
  \onecolumn
\section{Results of the abundance analysis}\label{sec:abu_app}
\subsection{Uncertainties}\label{ssec:uncert}
\begin{figure}[!ht]
\includegraphics[width=0.5\columnwidth,angle=0]{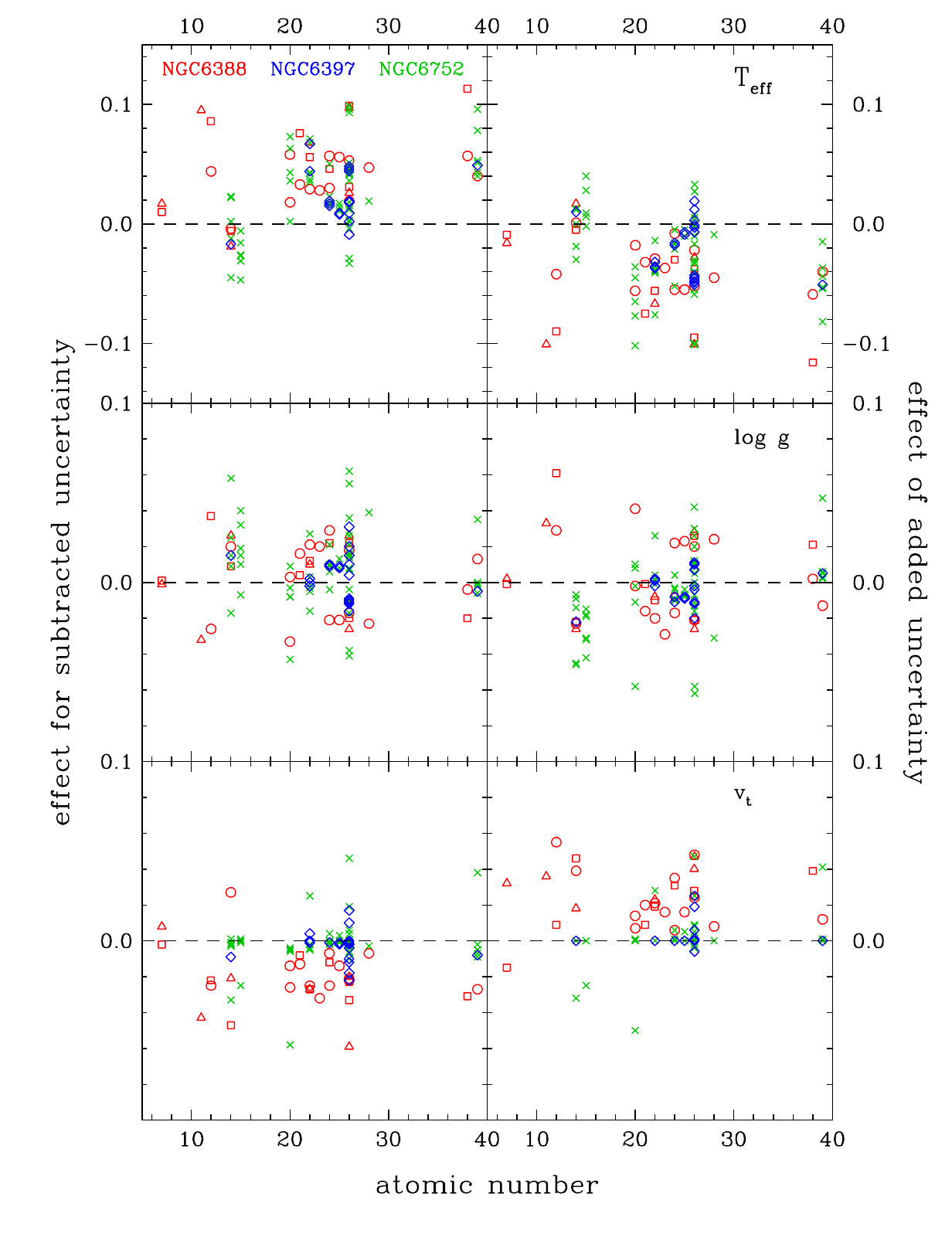}
\caption{Offsets in abundances determined by applying
    the estimated uncertainties (one at a time)
  for effective temperature (top row), surface gravity
  (middle row), and microturbulent velocity (bottom
    row) to the corresponding parameters determined by GALA
  and then rerunning GALA
  with fixed parameters. The left column shows the effect of subtracting the uncertainties, the right one the effect of adding the uncertainties.
 Red open symbols
  (circle, square, triangle) mark the stars in NGC\,6388, blue
  lozenges the ones in NGC\,6397, and green crosses the ones in
  NGC\,6752.}
\label{fig:abu_uncert}
 \end{figure}

To determine uncertainties in the abundances caused by uncertainties in the atmospheric parameter, I used the uncertainties listed below. 
The resulting offsets in abundances are shown in Fig.\,\ref{fig:abu_uncert}.

For NGC\,6388 I used the effective temperature and surface gravity
estimates marked in bold font in Table\,\ref{ngc6388_photpar} as
starting values. The atmospheric parameters and abundances
  resulting from the GALA analysis are listed in
Table\,\ref{tab:abu_photpar}. I used uncertainties in effective
temperature as follows: 50\,K for 1233, 250\,K for 4113, 120\,K for
5235, and 150\,K for 7788. For the surface gravity I used an
uncertainty of 0.1\,dex for all stars except 4113 (0.2\,dex).

\FloatBarrier

For the stars in NGC\,6397 and NGC\,6752, I used the atmospheric
parameters derived from spectral line fitting as starting values.
The results of the GALA analysis are listed in
Tables\,\ref{tab:abu_ngc6752_2000}, \ref{tab:abu_ngc6752_69D220} (both
for NGC\,6752), and \ref{tab:abu_ngc6397} (NGC\,6397).  For stars in
NGC\,6752 and NGC\,6397 observed with program 69.D-0220 I used an
uncertainty of 90\,K in effective temperature and of 0.04\,dex in
surface gravity. For the hotter stars in NGC\,6752 observed with
program 65.L-0233 I used uncertainties of 210\,K and 0.1\,dex in
effective temperature and surface gravity, respectively.
For all stars, I used an uncertainty of 0.2\,km\,s$^{-1}$ for the
 microturbulence.

\subsection{Abundances}\label{ssec:abu}
In this section, I list the abundances for the blue HB stars in NGC\,6388 (Table\,\ref{tab:abu_photpar}), NGC\,6397 (Table\,\ref{tab:abu_ngc6397}), and NGC\,6752 (Tables\,\ref{tab:abu_ngc6752_2000} and \ref{tab:abu_ngc6752_69D220}) derived as described in Sect.\,\ref{sec:abu}.
 \FloatBarrier

\begin{table*}
  \caption[]{Chemical abundances derived for NGC\,6388 stars with GALA using the atmospheric parameters derived from the HUGS photometry (Table\,\ref{ngc6388_photpar}) as starting values. Values in italics were obtained with errors of 15\% of the equivalent width. The references for the line parameters used for the analysis can be found in Table\,\ref{tab:vald_bib}.}\label{tab:abu_photpar}.
  \begin{tabular}{llrlrlrlr}
    \hline
    \hline
    & \multicolumn{2}{c}{1233} & \multicolumn{2}{c}{4113} & \multicolumn{2}{c}{5235} & \multicolumn{2}{c}{7788}\\
          $T_{eff}$ [K] & \multicolumn{2}{c}{8200} & \multicolumn{2}{c}{7100} & \multicolumn{2}{c}{9700} & \multicolumn{2}{c}{10250}\\
 $\log g$ [cm s$^{-2}$] & \multicolumn{2}{c}{2.85} & \multicolumn{2}{c}{2.95} & \multicolumn{2}{c}{3.70} & \multicolumn{2}{c}{3.65}\\
$v_{turb}$ [km s$^{-1}$] & \multicolumn{2}{c}{1.7} & \multicolumn{2}{c}{1.4} & \multicolumn{2}{c}{1.7} & \multicolumn{2}{c}{0.8}\\
    ion & [M/H] & n$_{lines}$ & [M/H] & n$_{lines}$ & [M/H] & n$_{lines}$ & [M/H] & n$_{lines}$ \\
\hline
 \hspace*{0pt}[Fe\,{\sc i}/H]  & $-$0.37$\pm$0.09 & 110 & $-$0.44$\pm$0.17 & 34 & $-$0.06$\pm$0.11 & 41 & $-$0.21$\pm$0.18 & 12\\
 \hspace*{0pt}[Fe\,{\sc ii}/H] & $-$0.37$\pm$0.12 &  44 & $-$0.43$\pm$0.10 &  7 & $-$0.03$\pm$0.13 & 41 & $-$0.23$\pm$0.15 & 16\\[1mm]
\hline
    ion & [M/Fe] & n$_{lines}$ & [M/Fe] & n$_{lines}$ & [M/Fe] & n$_{lines}$ & [M/Fe] & n$_{lines}$ \\
    \hline
\multicolumn{9}{c}{$\alpha$ elements}\\[1mm]
\hline
 \hspace*{0pt}[Si\,{\sc ii}/Fe] & $+$0.10$\pm$0.12 &  8 &         &    & $-$0.19$\pm$0.16 &  8 & $+$0.28$\pm$0.12 &  6\\
 \hspace*{0pt}[Ca\,{\sc i}/Fe]  & $+$0.16$\pm$0.14 & 15 &             &    &               &    &            &   \\
 \hspace*{0pt}[Ca\,{\sc ii}/Fe] & $+$0.01$\pm$0.07 &  5 &             &    &               &    &            &   \\
 \hspace*{0pt}[Ti\,{\sc ii}/Fe] & $-$0.01$\pm$0.07 & 37 &             &    & $-$0.03$\pm$0.10 & 26 & $-$0.13$\pm$0.05 &  8\\
\hline
\multicolumn{9}{c}{iron-peak elements}\\[1mm]
\hline
 \hspace*{0pt}[Sc\,{\sc ii}/Fe] & $-$0.14$\pm$0.08 &  6 &               &    & $-$0.42$\pm$0.23             &  3 &           &   \\
 \hspace*{0pt}[V\,{\sc ii}/Fe]  & $+$0.03$\pm$0.05 &  5 &               &    & {\it $+$0.01}$\pm${\it 0.10} & {\it 4} &      &   \\
 \hspace*{0pt}[Cr\,{\sc i}/Fe]  & $-$0.09$\pm$0.07 &  5 &               &    &                           &    &           &  \\ 
 \hspace*{0pt}[Cr\,{\sc ii}/Fe] & $-$0.01$\pm$0.07 & 21 & $-$0.38$\pm$0.14 &  5 & $-$0.08$\pm$0.04             &  7 &           &   \\
 \hspace*{0pt}[Mn\,{\sc i}/Fe]  & $-$0.16$\pm$0.18 &  4 &               &    &                           &    &           &   \\
 \hspace*{0pt}[Ni\,{\sc i}/Fe]  & $+$0.06$\pm$0.07 &  3 & $-$0.51$\pm$0.13 &  4 &                           &    &           & \\
\hline
\multicolumn{9}{c}{other elements}\\[1mm]
\hline
 \hspace*{0pt}[C\,{\sc i}/Fe] & {\it $-$0.61}$\pm${\it 0.18}   & {\it 3} &               &     &                           &         &                            &   \\
 \hspace*{0pt}[N\,{\sc i}/Fe] & {\it $+$1.47}$\pm${\it 0.08}   & {\it 7} &               &     & {\it $+$1.34}$\pm${\it 0.12} & {\it 5} & {\it $+$1.46}$\pm${\it 0.07} & {\it 3} \\
 \hspace*{0pt}[O\,{\sc i}/Fe] & {\it $-$0.37}$\pm${\it 0.02}   & {\it 2} &               &     &                           &         & {\it $-$0.35}$\pm${\it 0.02} & {\it 3} \\
 \hspace*{0pt}[Na\,{\sc i}/Fe] & {\it $+$0.85}$\pm${\it }0.17  &   {\it 2} &               &     & {\it $+$0.97}$\pm${\it 0.13} &  {\it 2}  & $+$1.02$\pm$0.06             & 2\\
 \hspace*{0pt}[Mg\,{\sc i}/Fe] & $+$0.12$\pm$0.18              &       7 & $+$0.08$\pm$0.13 &  2  & $+$0.08$\pm$0.12          &       5 & {\it $+$0.30}$\pm${\it 0.01} & {\it 2}\\
 \hspace*{0pt}[Al\,{\sc i}/Fe] &                               &         &               &     &                           &         & {\it $-$0.08}$\pm${\it 0.03} & {\it 2}\\
 \hspace*{0pt}[Sr\,{\sc ii}/Fe] & $+$0.16$\pm$0.06             &       2 &               &     & $-$0.54$\pm$0.02             &  2 &                           &   \\
 \hspace*{0pt}[Y\,{\sc ii}/Fe] & {\it $-$0.21}$\pm${\it 0.11} & {\it 5} &               &     &            &    &            &   \\
 \hspace*{0pt}[La\,{\sc ii}/Fe] & $+$1.90$\pm$0.23             &  2      &               &     &            &    &            &   \\
\hline
  \end{tabular}
  \end{table*}

\begin{table*}
  \caption[]{Chemical abundances derived for NGC\,6752 stars (65.L-0233(A)) with GALA using atmospheric parameters from spectral line profile fitting (Table\,\ref{tab:metal_poor_param_diff}) as starting values. Values in italics were obtained with errors of 15\% of the equivalent width.  The references for the line parameters used for the analysis can be found in Tables\,\ref{tab:vald_bib} and \ref{tab:vald_bib2}.}\label{tab:abu_ngc6752_2000}. 
  \begin{tabular}{llrlrlrlrlr}
    \hline
    \hline
    & \multicolumn{2}{c}{B1509} & \multicolumn{2}{c}{B2099} & \multicolumn{2}{c}{B2698} & \multicolumn{2}{c}{B4172} & \multicolumn{2}{c}{B4598}\\
    $T_{eff}$ [K] & \multicolumn{2}{c}{16450} & \multicolumn{2}{c}{17350} & \multicolumn{2}{c}{14000}
    & \multicolumn{2}{c}{12600} & \multicolumn{2}{c}{13250}\\
    $\log g$ [cm s$^{-2}$]  & \multicolumn{2}{c}{4.55} & \multicolumn{2}{c}{4.45} & \multicolumn{2}{c}{4.30}
    & \multicolumn{2}{c}{4.10} & \multicolumn{2}{c}{4.20}\\
    $v_{turb}$ [km s$^{-1}$]  & \multicolumn{2}{c}{0.0} & \multicolumn{2}{c}{0.0} & \multicolumn{2}{c}{0.0}
    & \multicolumn{2}{c}{0.0} & \multicolumn{2}{c}{0.0}\\
    ion & [M/H] &  n$_{lin}$ & [M/H] & n$_{lin}$ & [M/H] &  n$_{lin}$      & [M/H] &  n$_{lin}$ & [M/H] &  n$_{lin}$ \\
\hline
\hspace*{0pt}[Fe\,{\sc i}/H]   &                  &    &                  &    & $+$0.54$\pm$0.06 &  4 & $+$0.91$\pm$0.11 &  30 & $+$1.19$\pm$0.12 & 30 \\
\hspace*{0pt}[Fe\,{\sc ii}/H]  & $+$0.97$\pm$0.11 & 70 & $+$0.62$\pm$0.18 & 20 & $+$0.54$\pm$0.11 & 62 & $+$0.92$\pm$0.10 & 107 & $+$1.17$\pm$0.12 &118 \\
\hspace*{0pt}[Fe\,{\sc iii}/H] & $+$0.97$\pm$0.12 &  6 & $+$0.58$\pm$0.02 &  2 &               &                    &     &               &  &  \\
\hline
\multicolumn{11}{c}{$\alpha$ elements}\\[1mm]
\hline
\hspace*{0pt}[Si\,{\sc ii}/H]   & {\it $+$0.25$\pm$0.11} & {\it 6} & $-$0.30$\pm$0.03 & 2 & $-$1.04$\pm$0.03 & 4 &       &  &           &  \\
\hspace*{0pt}[Ca\,{\sc ii}/H]   &       &  &       &  &       &                & $-$0.27$\pm$0.21 & 2 & $+$0.22$\pm$0.20 & 2\\
\hspace*{0pt}[Ti\,{\sc ii}/H]   &       &  &       &  & {\it $+$0.16$\pm$0.05} & {\it 3}  & $+$0.52$\pm$0.09 & 19 & $+$0.21$\pm$0.09 & 11\\
\hline
\multicolumn{11}{c}{iron-peak elements}\\[1mm]
\hline
\hspace*{0pt}[Cr\,{\sc ii}/H]   & {\it $+$0.82$\pm$0.09} & {\it 5} &       &  &       &                & $-$0.04$\pm$0.20 & 8 &          &  \\
\hspace*{0pt}[Ni\,{\sc ii}/H]   &       &  & {\it $+$1.51$\pm$0.18} & {\it 5} &     &                &            &   & $+$0.30$\pm$0.06 & 3\\ 
\hline
\multicolumn{11}{c}{other elements}\\[1mm]
\hline
\hspace*{0pt}[P\,{\sc ii}/H]  & $+$1.95$\pm$0.12 & 8 &       &  & {\it $+$0.83$\pm$0.03} & {\it 3} & $+$1.18$\pm$0.10 & 5 & {\it $+$1.66$\pm$0.13} & {\it 8}\\
\hspace*{0pt}[Y\,{\sc ii}/H]  &       &  &       &  &       &  & $+$2.41$\pm$0.08 & 3 & $+$2.92$\pm$0.10 & 4 \\
\hline
  \end{tabular}
  \end{table*}

\begin{table*}
  \caption[]{Chemical abundances derived for NGC\,6752 stars
    (69.D-0220(A)) with GALA using atmospheric parameters from
    spectral line profile fitting as starting values
    (Table\,\ref{tab:metal_poor_param_diff}). The references for the
    line parameters used for the analysis can be found in
    Table\,\ref{tab:vald_bib}.}
  \label{tab:abu_ngc6752_69D220} 
  \begin{tabular}{llrlrlrlrlr}
    \hline
    \hline
    & \multicolumn{2}{c}{B652} & \multicolumn{2}{c}{B2151} & \multicolumn{2}{c}{B2206} & \multicolumn{2}{c}{B2649}
    & \multicolumn{2}{c}{B3243} \\
    $T_{eff}$ [K] &  \multicolumn{2}{c}{12100} & \multicolumn{2}{c}{11200} & \multicolumn{2}{c}{11050} & \multicolumn{2}{c}{10950}
    & \multicolumn{2}{c}{10450} \\
    $\log g$ [cm s$^{-2}$] & \multicolumn{2}{c}{3.90} & \multicolumn{2}{c}{3.45} & \multicolumn{2}{c}{3.40} & \multicolumn{2}{c}{3.35}
    & \multicolumn{2}{c}{3.25} \\
    $v_{turb}$ [km s$^{-1}$] & \multicolumn{2}{c}{0.0} & \multicolumn{2}{c}{0.0} & \multicolumn{2}{c}{0.0} & \multicolumn{2}{c}{0.0}
    & \multicolumn{2}{c}{0.0} \\
    ion & [M/H] & n$_{lin}$ & [M/H] &  n$_{lin}$ & [M/H] & n$_{lin}$ & [M/H] &  n$_{lin}$
    & $\log \epsilon $ \\
\hline
\hspace*{0pt}[Fe\,{\sc i}/H] & $+$0.72$\pm$0.14 & 15 & $+$0.35$\pm$0.05 & 29 & $+$0.36$\pm$0.06 & 27  & $-$0.45$\pm$0.05 & 10 
              & $+$0.30$\pm$0.05 & 31  \\
\hspace*{0pt}[Fe\,{\sc ii}/H]  & $+$0.72$\pm$0.11 & 149 & $+$0.35$\pm$0.10 & 182 & $+$0.35$\pm$0.10 & 167 & $-$0.47$\pm$0.10 & 39 
              & $+$0.20$\pm$0.08 & 132 \\
\hline
\multicolumn{11}{c}{$\alpha$ elements}\\[1mm]
\hline
\hspace*{0pt}[Si\,{\sc ii}/H] &      &  &       &  &       &  & $-$1.12$\pm$0.22 & 10 
              & $-$2.18$\pm$0.21 & 4  \\
\hspace*{0pt}[Ca\,{\sc ii}/H]  &               &   & $-$0.63$\pm$0.23 & 3 &        &   & $-$0.98$\pm$0.28 & 3 
              & $-$0.90$\pm$0.13 & 3 \\
\hspace*{0pt}[Ti\,{\sc ii}/H]  & {\it $-$0.07$\pm$0.11} & {\it 11} & $+$0.06$\pm$0.04 & 24 & $+$0.06$\pm$0.06 & 24 & $+$0.55$\pm$0.06 & 49 
              & $+$0.02$\pm$0.05 & 29  \\
\hline
\multicolumn{11}{c}{iron-peak elements}\\[1mm]
\hline
\hspace*{0pt}[Cr\,{\sc ii}/H] & {\it $-$0.31$\pm$0.05} & {\it 2} &      &   & $-$0.15$\pm$0.04 & 7 & $-$0.15$\pm$0.04 & 7 
              &  {\it $-$0.28$\pm$0.04} & {\it 6}  \\
\hspace*{0pt}[Mn\,{\sc ii}/H] & {\it $+$0.94$\pm$0.18} & {\it 16} &      &  & $+$0.24$\pm$0.07 & 2 & $+$0.90$\pm$0.09 & 11 
              & {\it $+$0.29$\pm$0.01} & {\it 2}  \\
\hline
\multicolumn{11}{c}{other elements}\\[1mm]
\hline
\hspace*{0pt}[P\,{\sc ii}/H] & {\it $+$1.50$\pm$0.14} & {\it 6} & $+$0.88$\pm$0.06 & 2 & {\it $+$0.87$\pm$0.03} & {\it 2}  &       & 
              &  &   \\
\hspace*{0pt}[Y\,{\sc ii}/H] & $+$2.18$\pm$0.04 & 2 & {\it $+$1.22$\pm$0.05} & {\it 3} & $+$1.46$\pm$0.07 & 5 & $+$1.52$\pm$0.09 & 6
              &  &   \\
\hline
  \end{tabular}
  \end{table*}

\begin{table*}
  \caption[]{The chemical abundances derived for NGC\,6397 stars (69.D-0220(B)) with GALA using atmospheric parameters from spectral line profile fitting as starting values (Table\,\ref{tab:metal_poor_param_diff}).  The references for the line parameters used for the analysis can be found in Table\,\ref{tab:vald_bib}.}\label{tab:abu_ngc6397}
  \begin{tabular}{llrlrlrlr}
    \hline
    \hline
    & \multicolumn{2}{c}{T170} & \multicolumn{2}{c}{T172} & \multicolumn{2}{c}{T179} & \multicolumn{2}{c}{T183}\\
    $T_{eff}$ [K] &  \multicolumn{2}{c}{11550} & \multicolumn{2}{c}{11100} & \multicolumn{2}{c}{10800} & \multicolumn{2}{c}{11300}\\
    $\log g$ [cm s$^{-2}$] & \multicolumn{2}{c}{3.80} & \multicolumn{2}{c}{3.75} & \multicolumn{2}{c}{3.40} & \multicolumn{2}{c}{3.30}\\
    $v_{turb}$ [km s$^{-1}$] & \multicolumn{2}{c}{0.7} & \multicolumn{2}{c}{0.2} & \multicolumn{2}{c}{0.3} & \multicolumn{2}{c}{0.0}\\
    \hline
    ion & [M/H] & n$_{lin}$ & [M/H] &  n$_{lin}$ & [M/H] & n$_{lin}$ & [M/H] &  n$_{lin}$\\
\hline
\hspace*{0pt}[Fe\,{\sc i}/H]   & $-$0.24$\pm$0.06 & 10 & $+$0.08$\pm$0.08 & 10 & $-$0.18$\pm$0.07 & 9  & $+$0.25$\pm$0.10 & 6\\
\hspace*{0pt}[Fe\,{\sc ii}/H]  & $-$0.22$\pm$0.12 & 37 & $+$0.07$\pm$0.14 & 46 & $-$0.17$\pm$0.09 & 50 & $+$0.27$\pm$0.15 & 47\\
\hline
\multicolumn{9}{c}{$\alpha$ elements}\\[1mm]
\hline
\hspace*{0pt}[Si\,{\sc ii}/H] &               &   & {\it $-$1.79$\pm$0.08} &{\it 4} & $-$1.67$\pm$0.06 &  5 &               &\\
\hspace*{0pt}[Ti\,{\sc ii}/H] &               &   &               &    &               &    & {\it $+$0.45$\pm$0.06} & {\it 17}\\
\hline
\multicolumn{9}{c}{iron-peak elements}\\[1mm]
\hline
\hspace*{0pt}[Mn\,{\sc ii}/H] &               &   &               &    & {\it $+$0.36$\pm$0.03} & {\it 2} &              &\\
\hline
\multicolumn{9}{c}{}\\
\hline
\hline
 & \multicolumn{2}{c}{T185} & \multicolumn{2}{c}{T186} & \multicolumn{2}{c}{T191} & \multicolumn{2}{c}{T193}\\
    $T_{eff}$ [K] & \multicolumn{2}{c}{11050} & \multicolumn{2}{c}{11250} & \multicolumn{2}{c}{11150} & \multicolumn{2}{c}{11550}\\
    $\log g$ [cm s$^{-2}$]  & \multicolumn{2}{c}{3.50} & \multicolumn{2}{c}{3.60} & \multicolumn{2}{c}{3.60} & \multicolumn{2}{c}{3.60}\\
    $v_{turb}$ [km s$^{-1}$] & \multicolumn{2}{c}{0.0} & \multicolumn{2}{c}{0.0} & \multicolumn{2}{c}{0.0} & \multicolumn{2}{c}{0.0}\\
    \hline
    ion  & [M/H] & n$_{lin}$ & [M/H] &  n$_{lin}$ & [M/H] & n$_{lin}$ & [M/H] &  n$_{lin}$ \\
    \hline
\hspace*{0pt}[Fe\,{\sc i}/H]   & $-$0.26$\pm$0.05 & 8  & $+$0.09$\pm$0.09 & 12 & $-$0.21$\pm$0.12 & 3  & $+$0.25$\pm$0.10 & 13\\
\hspace*{0pt}[Fe\,{\sc ii}/H]  & $-$0.28$\pm$0.06 & 29 & $+$0.10$\pm$0.13 & 79 & $-$0.32$\pm$0.10 & 19 & $+$0.27$\pm$0.11 & 96 \\
\hline
\multicolumn{9}{c}{$\alpha$ elements}\\[1mm]
\hline
\hspace*{0pt}[Ti\,{\sc ii}/H]  &               &   & $+$0.67$\pm$0.08 & 39 & $+$1.28$\pm$0.09 & 19 & $+$0.34$\pm$0.06 & 19\\
\hline
\multicolumn{9}{c}{iron-peak elements}\\[1mm]
\hline
\hspace*{0pt}[Cr\,{\sc ii}/H] & {\it $-$0.45$\pm$0.04} & {\it 3} & $+$0.08$\pm$0.09 &  9 & {\it $+$0.04$\pm$0.01} & {\it 2} & $-$0.13$\pm$0.06 & 4\\
\hspace*{0pt}[Mn\,{\sc ii}/H] &               &   & $+$1.10$\pm$0.15 & 10 & $+$1.03$\pm$0.23 & 8 & $+$1.18$\pm$0.10 & 11\\
\hline
\multicolumn{9}{c}{other elements}\\[1mm]
\hline
\hspace*{0pt}[Y\,{\sc ii}/H] &                &   & $+$1.71$\pm$0.03 & 3 &               &   & {\it $+$1.67$\pm$0.03} & {\it 2}\\
\hline
  \end{tabular}
\end{table*}

\section{References for VALD line lists}\label{sec:vald_ref}
In addition to the references listed below the following publications have been used for isotopic scaling: \citet[Ti \,{\sc ii}]{NRLS}, \citet[Ca\,{\sc ii}]{MYWVS}, \citet[Ni\,{\sc i}]{K08}
\begin{table*}[!h]
  \caption[]{References for the lines lists obtained from VALD for the abundance analysis of the stars in NGC\,6397 and NGC\,6752.}\label{tab:vald_bib}
  \begin{tabular}{l|p{4cm}|p{7cm}|p{4cm}}
    \hline
    \hline
    ion & wavelength & transition & broadening \\
               &     & probability & parameters \\
    \hline
      \multicolumn{4}{c}{65.L-0233 (NGC\,6752)}\\
    \hline
     Fe\,{\sc i} & \cite{K14} & \cite{BK}, \cite{BWL}, \cite{FMW}, \cite{K14} & \cite{BPM}, \cite{K14}\\
     Fe\,{\sc ii} & \cite{BSScor}, \cite{K13} & \cite{BGHR}, \cite{BSScor}, \cite{K13}, \cite{KK}, \cite{PGHcor}, \cite{RU}, \cite{T83av} & \cite{K13}, \cite{BA-J}\\
     Si\,{\sc ii} & \cite{K14} & \cite{MER}, \cite{NIST10}, \cite{S-G}, \cite{Si2-av1} & \cite{K14}\\
     P\,{\sc ii} & \cite{K12} & \cite{K12} & \cite{K12}\\
     Ca\,{\sc ii}$^\ast$ & \cite{T} & \cite{T} & \cite{BPM}, \cite{T}\\ 
     Ti\,{\sc ii}$^\ast$ & \cite{K16}, \cite{Sal12a} & \cite{PTP}, \cite{RHL}, \cite{WLSC} & \cite{K16}\\ 
     Cr\,{\sc ii} & \cite{SNave} & \cite{LSNE} & \cite{BA-J}, \cite{K16}\\
     Ni\,{\sc ii} & \cite{K03} & \cite{K03} & \cite{K03}\\
     Y\,{\sc ii} & \cite{K11} & \cite{HLGBW} & \cite{K11}\\
    \hline
      \multicolumn{4}{c}{69.D-0220 (NGC\,6397, NGC\,6752)}\\
    \hline
     Fe\,{\sc i} & \cite{BWL}, \cite{K14} & \cite{FMW}, \cite{BWL} & \cite{BPM}, \cite{K14}\\
     Fe\,{\sc ii} & \cite{BSScor}, \cite{K13} & \cite{B}, \cite{BGHR}, \cite{BSScor}, \cite{HLGN}, \cite{K13}, \cite{KK}, \cite{PGHcor}, \cite{RU}, \cite{T83av} & \cite{K13}, \cite{BA-J}\\
     Si\,{\sc ii} & \cite{K14} & \cite{K14}, \cite{MER}, \cite{S-G}, \cite{Si2-av1} & \cite{K14}\\
     P\,{\sc ii} & \cite{K12} & \cite{K12} & \cite{K12}\\
     Ca\,{\sc ii}$^\ast$ & \cite{K10} & \cite{K10} & \cite{BPM}, \cite{K10}, \cite{T}\\
     Ti\,{\sc ii} & \cite{K16}, \cite{Sal12a} & \cite{PTP}, \cite{RHL}, \cite{WLSC} & \cite{K16}\\
     Cr\,{\sc ii} & \cite{K16}, \cite{SNave} & \cite{LSNE}, \cite{RU} & \cite{BA-J}, \cite{K16}\\
     Mn\,{\sc ii} & \cite{K09} & \cite{K09}, \cite{KSG} & \cite{K09}\\
     Y\,{\sc ii} & \cite{K11} & \cite{HLGBW} & \cite{K11}\\
     \hline
     \end{tabular}
\end{table*}
\FloatBarrier
\begin{table*}[!h]
  \caption[]{This table provides the references for the lines lists obtained from VALD for the abundance analysis of the stars in NGC\,6388.}\label{tab:vald_bib2}
  \begin{tabular}{l|p{4cm}|p{7cm}|p{4cm}}
    \hline
    \hline
    ion & wavelength & transition & broadening \\
               &     & probability & parameters \\
    \hline  
    \multicolumn{4}{c}{69.D-0231 (NGC\,6388)}\\
    \hline
     Fe\,{\sc i} & \cite{BK}, \cite{BWL}, \cite{K14} & \cite{BK}, \cite{BKK}, \cite{BWL}, \cite{FMW}, \cite{K14} & \cite{BPM}, \cite{K14}\\
     Fe\,{\sc ii} & \cite{BSScor}, \cite{K13} & \cite{B}, \cite{BGHR}, \cite{BSScor}, \cite{HLGN}, \cite{K13}, \cite{KK}, \cite{PGHcor}, \cite{RU}, \cite{T83av} & \cite{K13}, \cite{BA-J}\\
     C\,{\sc i} & \cite{NIST10} & \cite{NIST10} & \cite{NIST10}\\
     N\,{\sc i} & \cite{NIST10} & \cite{NIST10} & \cite{BPM}, \cite{NIST10}\\
     O\,{\sc i} & \cite{NIST10} & \cite{NIST10} & \cite{BPM}\\
     Na\,{\sc i} & \cite{NIST10} & \cite{NIST10} & \cite{BPM}, \cite{WSG}\\
     Mg\,{\sc i} & \cite{NIST10} & \cite{ATJL}, \cite{BPM}, \cite{NIST10} & \cite{AZS}, \cite{BPM}, \cite{LZ}, \cite{KP}\\
     Al\,{\sc i} & \cite{SLa} & \cite{SLa} & \cite{BPM}, \cite{SLa}\\
     Si\,{\sc ii} & \cite{K14}, \cite{K17} & \cite{BBC}, \cite{K17}, \cite{S-G}, \cite{Si2-av1} & \cite{BPM}, \cite{K14}, \cite{K17}\\
     Ca\,{\sc i} & \cite{K07}, \cite{S}, \cite{SG}, \cite{SN}, \cite{SR} & \cite{K07}, \cite{S}, \cite{SG}, \cite{SN}, \cite{SR} & \cite{BPM}, \cite{K07}, \cite{S}, \cite{SG}, \cite{SN}, \cite{SR}\\
     Ca\,{\sc ii} & \cite{TB} & \cite{TB} & \cite{TB}\\
     Sc\,{\sc ii} & \cite{K09}, \cite{NIST22} & \cite{LHSNWC} & \cite{K09}\\
     Ti\,{\sc ii}$^\ast$ & \cite{K16}, \cite{Sal12a} & \cite{PTP}, \cite{RHL}, \cite{WLSC} & \cite{K16}\\ 
     V\,{\sc ii} & \cite{TPSb} & \cite{WLDSC} & \cite{K10}\\
     Cr\,{\sc i} & \cite{NIST22} & \cite{SLS} & \cite{BPM}, \cite{K16}\\
     Cr\,{\sc ii} & \cite{SNave} & \cite{LSNE} & \cite{BA-J}, \cite{K16}\\
     Mn\,{\sc i} & \cite{SugCor} & \cite{DLSSC} & \cite{BPM}, \cite{K07}\\
     Ni\,{\sc i}$^\ast$ & \cite{LBT} & \cite{FMW}, \cite{K08}, \cite{WLSCow} & \cite{BPM}, \cite{K08}\\ 
     Sr\,{\sc ii} & \cite{K17} & \cite{K17} & \cite{BPM}, \cite{K17}\\
     Y\,{\sc ii} & \cite{K11} & \cite{HLGBW} & \cite{K11}\\
     La\,{\sc ii} & \cite{CB}, \cite{LBS} & \cite{CB}, \cite{LBS} & \cite{CB}, \cite{LBS} \\
    \hline
  \end{tabular}
\end{table*}

\end{appendix}
\end{document}